\documentclass[12pt]{article}
\usepackage[margin=.5in]{geometry}
\usepackage{fullpage}
\usepackage[utf8]{inputenc}
\usepackage{graphicx}
\usepackage{subfigure}
\usepackage[colorlinks=true, citecolor=blue]{hyperref}
\usepackage{array}
\usepackage{ulem}
\usepackage{epstopdf}
\usepackage{bbm}
\usepackage{mathtools}
\usepackage{empheq}
\usepackage{amsfonts}
\usepackage{amsthm}
\usepackage{amsgen,amsmath,amstext,amsbsy,amsopn,amssymb,stmaryrd}
\usepackage{enumitem}
\SetSymbolFont{stmry}{bold}{U}{stmry}{m}{n}
\usepackage{comment}
\usepackage{natbib}
\usepackage{booktabs}
\usepackage{blkarray}
\usepackage{algorithm}
\usepackage{algpseudocode} 
\usepackage{float}
\usepackage{makecell}
\usepackage{url}
\usepackage{longtable}
\usepackage{tabularx}
\usepackage{multirow}
\usepackage[capitalize]{cleveref}
\usepackage[dvipsnames]{xcolor}
\usepackage{tikz}
\usetikzlibrary{arrows.meta, chains, positioning}
\usepackage{caption}
\usepackage{titling}
\usepackage{float}

\makeatletter
\def\@footnotecolor{red}

\define@key{Hyp}{footnotecolor}{%
	\HyColor@HyperrefColor{#1}\@footnotecolor%
}
\def\@footnotemark{%
	\leavevmode
	\ifhmode\edef\@x@sf{\the\spacefactor}\nobreak\fi
	\stepcounter{Hfootnote}%
	\global\let\Hy@saved@currentHref\@currentHref
	\hyper@makecurrent{Hfootnote}%
	\global\let\Hy@footnote@currentHref\@currentHref
	\global\let\@currentHref\Hy@saved@currentHref
	\hyper@linkstart{footnote}{\Hy@footnote@currentHref}%
	\@makefnmark
	\hyper@linkend
	\ifhmode\spacefactor\@x@sf\fi
	\relax
}%
\makeatother

\hypersetup{footnotecolor=black}

\newcolumntype{P}[1]{>{\centering\arraybackslash}p{#1}}

\ifpdf
\DeclareGraphicsExtensions{.eps,.pdf,.png,.jpg}
\else
\DeclareGraphicsExtensions{.eps}
\fi

\DeclareMathAlphabet\mathbfcal{OMS}{cmsy}{b}{n}

\newcommand{\be}{\begin{equation}}
	\newcommand{\ee}{\end{equation}}
\newcommand{\bea}{\begin{eqnarray}}
	\newcommand{\eea}{\end{eqnarray}}
\newcommand{\beas}{\begin{eqnarray*}}
	\newcommand{\eeas}{\end{eqnarray*}}

\newcommand{\argmin}{\mathop{\rm arg\min}}
\newcommand{\argmax}{\mathop{\rm arg\max}}

\renewcommand{\hat}{\widehat}

\newcommand{\comm}[1]{}

\makeatletter
\newcommand*{\rom}[1]{\expandafter\@slowromancap\romannumeral #1@}
\makeatother 

\allowdisplaybreaks

\title{Subtype-Aware Registration of Longitudinal Electronic Health Records}
\author{Xin Gai$^{1}$\thanks{Department of Biostatistics, Vanderbilt University}, ~ Shiyi Jiang$^{1}$\thanks{Department of Electrical \& Computer Engineering, Duke University}, ~ and ~ Anru R. Zhang$^{2}$\thanks{Department of Biostatistics \& Bioinformatics and Department of Computer Science, Duke University}}

\date{}

\begin{document}
	
	\maketitle
	
	\def\thefootnote{}\footnotetext{$^1$Equal contribution. $^2$Corresponding author. E-mail: \texttt{anru.zhang@duke.edu}}\def\thefootnote{\arabic{footnote}}
	
	\begin{abstract}
		Electronic Health Records (EHRs) contain extensive patient information that can inform downstream clinical decisions, such as mortality prediction, disease phenotyping, and disease onset prediction. A key challenge in EHR data analysis is the temporal gap between when a condition is first recorded and its actual onset time. Such timeline misalignment can lead to artificially distinct biomarker trends among patients with similar disease progression, undermining the reliability of downstream analyses and complicating tasks such as disease subtyping and outcome prediction. To address this challenge, we provide a subtype-aware timeline registration method that leverages data projection and discrete optimization to correct timeline misalignment. Through simulation and real-world data analyses, we demonstrate that the proposed method effectively aligns distorted observed records with the true disease progression patterns, enhancing subtyping clarity and improving performance in downstream clinical analyses.
	\end{abstract}

    \begin{sloppypar}
	\section{Introduction} \label{sec:intro}
	
		Electronic Health Records (EHR) are digitized collections of patient hospital information, encompassing data such as demographics, lab measurements, diagnoses, and medications. Data collected from intensive care unit (ICU) patients are of particular interest in clinical research due to their high frequency and timely collection. These data contain crucial information for executing timely disease interventions, managing patient care, and facilitating various downstream clinical decisions \cite{Shickel2017DeepEA, Yadav2017MiningEH, Shillan2019UseOM}. 
	
	However, a significant challenge in utilizing EHR data lies in the temporal gap between when a condition is first recorded and its actual onset time. For example, a patient with pneumonia might be admitted to the hospital at the congestion stage (stage one) \cite{Torres2021Pneumonia}, while another patient might delay hospital visitation until a later stage due to factors like insurance coverage or personal health beliefs \cite{Irlmeier2022Cox}, resulting in unobserved data for the initial stages of disease progression. Conventional analyses on EHR data often overlook this misalignment of timelines, potentially leading to biased modeling results. Therefore, it is crucial to address this issue by developing methods that can effectively align observed EHR timelines with the true underlying disease progression.
	
	\subsection{Registration in the Literature} \label{sec:related}
	
	This need for timeline alignment is closely related to the problem of curve registration in functional data analysis, which seeks to align curves so that key features are properly synchronized across subjects \cite{wang2016functional,marron2015functional,Ramsay2002CurveR}. In engineering, this problem is often referred to as time warping \cite{sakoe1978dynamic, wang1999synchronizing}. Early methods aligned curves to a single baseline, using approaches like self-modeling nonlinear regression \cite{lawton1971self, kneip1988convergence} or template selection through functional principal component analysis (FPCA) \cite{yao2005functional}, e.g.,  \cite{Kneip2008CombiningRA}. Subsequent work has explored a variety of alignment strategies. For instance, James \cite{james2007curve} proposed equating curve “moments” to capture local and global features, while Sangalli et al. \cite{sangalli2009case} introduced a similarity index to iteratively register functional data. Other studies have applied Fisher-Rao metric-based approaches \cite{srivastava2011registration, tucker2013generative, wu2014analysis}, computationally efficient techniques tailored for exponential family functional data \cite{wrobel2019registration,wrobel2018register,wrobel2021registr}, and pairwise warping strategies \cite{tang2008pairwise}. These approaches have been applied to a range of applications, including house price modeling \cite{peng2014time}, accelerometric activity data \cite{mcdonnell2022registration}, and speech analysis \cite{hadjipantelis2015unifying}. However, since EHR data are often highly heterogeneous, sparse, noisy, and irregular, traditional curve registration methods are often not directly applicable. 
	
	Furthermore, some studies have explored combining curve registration with clustering to account for the latent subgroups among subjects. Liu et al. \cite{liu2009simultaneous} used an EM algorithm with B-splines, while Wu et al. \cite{wu2016bayesian} employed a Bayesian framework that accommodates flexible warping functions. Yet, these methods often rely on specific data distributions that may not align with the complexities of EHR datasets. 
	
	To the best of our knowledge, Jiang et al. \cite{jiang2023timeline} made a first attempt to address EHR-specific registration. They proposed aligning EHR trajectories by estimating time shifts to a common disease progression template. They assumed that all patients with a disease of interest share a common disease progression template and proposed a registration method using alternating optimization to align observed EHR data to this shared template. Despite its novelty, this approach has several limitations: (\romannumeral 1) the reliance on linear interpolation and extrapolation for imputing missing data can lead to inaccurate estimations, particularly in sparsely observed EHR datasets; (\romannumeral 2) the high computational burden of timeline registration for imputed curves becomes impractical for large patient cohorts; and (\romannumeral 3) the assumption of a single progression template fails to account for the significant heterogeneity in diseases, where patients may follow distinct progression patterns. 
	
	In summary, while various registration and clustering methods exist, they remain constrained by assumptions unsuitable for EHR data’s sporadic and heterogeneous nature. These challenges necessitate the development of a new timeline registration method that effectively addresses the distinct complexities of EHR datasets, including data sparsity, diverse disease subtypes, and large patient populations.
	
	\subsection{Our Contributions} \label{sec:contribution}
	
	In this paper, we develop a subtype-aware timeline registration framework for longitudinal EHR data, with a focus on ICU AKI trajectories. Our registration framework simultaneously aligns patient timelines and discovers latent progression subtypes. By allowing different clusters of patients to follow distinct temporal patterns, our method moves beyond single-template registration and better reflects the clinical heterogeneity of ICU populations. In the proposed registration framework, we design a three-step algorithm that first embeds irregular trajectories into a low-dimensional B-spline representation, then iteratively alternates between clustering and optimizing subject-specific time shifts, and finally refines assignments while trimming potential outliers. This design yields substantial computational savings and robustness to noisy measurements and aberrant trajectories, making it suitable for large-scale EHR databases.
		
	Through extensive simulations, we show that the proposed method can recover latent time shifts with small error across a wide range of trajectory shapes, mean structures, and progression speeds, with runtimes under a few minutes. In real data from the MIMIC-IV ICU AKI cohort, our registration improves phenotype separation, yields clinically interpretable AKI subtypes, and increases severe AKI prediction performance by 3–5\% compared to unregistered data and an existing EHR registration baseline. The corresponding R package implementing the proposed registration method is publicly available at \url{https://github.com/LeoGai/subtypeReg}, and the source code used to reproduce the analyses in this paper can be found at \url{https://github.com/LeoGai/EHR-registration-subtype}.
	
	\subsection{Organization of the paper}
	The rest of the paper is organized as follows. In Section~\ref{sec:data_problem}, we describe the ICU EHR database, the clinical question, and our formal problem setup. Section~\ref{sec:methods} presents the proposed subtype-aware registration algorithm and its computational properties. Section~\ref{sec:real-data} presents a comprehensive analysis of AKI trajectories in the MIMIC-IV ICU cohort, including unsupervised phenotyping, clinical evaluation, and supervised prediction. We conclude with discussions of limitations and future directions in Section \ref{sec:discussions}. Supplementary Section~A reports extensive simulation results, and Supplementary Section~C provides an analysis of the computational and space complexities.

	\section{EHR Database and Problem Formulation} \label{sec:data_problem}
	
	\subsection{ICU EHR Database and Clinical question}
	
	We analyze ICU EHR data from the MIMIC-IV database, which contains detailed clinical information on over 40{,}000 ICU stays at the Beth Israel Deaconess Medical Center between 2008 and 2019 \cite{Goldberger2000PhysionetCO,Johnson2023MIMICIVAF,johnson2023mimicivdemo}. For this study, we focus on patients with acute kidney injury (AKI), a common clinical condition characterized by an abrupt decline in kidney function, resulting from various causes such as decreased renal perfusion, direct nephron injury, or urinary tract obstruction \cite{Makris2016AcuteKI}. AKI can develop rapidly within hours or days and is associated with significant morbidity and high mortality rates \cite{Rewa2014AcuteKI}. This syndrome is highly heterogeneous, and the complexity of its pathophysiology poses significant challenges in developing effective patient treatment plans \cite{Rewa2014AcuteKI}.
	
	We identify ICU stays with ICD codes indicating acute kidney failure and text descriptions containing the phrase ``acute kidney injury.'' We exclude patients with burns, end-stage renal disease, or renal dialysis during their ICU stay, as well as those with an estimated glomerular filtration rate (eGFR) below 15 mL/min/1.73 m$^2$ to avoid including end-stage renal disease \cite{song2020cross}. For data quality, we remove ICU stays with fewer than 15 recorded serum creatinine (SCr) measurements in the first 21 days after ICU admission, and those with two or fewer SCr measurements in each of the intervals 0–7, 7–14, and 14–21 days.
	
	Serum creatinine is the primary laboratory marker used to diagnose and stage AKI and to guide treatment decisions \cite{Khwaja2012KDIGOCP}. However, the first recorded SCr in the ICU does not generally correspond to the same disease stage across patients. Some patients are admitted early in their AKI course, whereas others arrive after several days of unobserved deterioration. Our central clinical questions are:
	
	\begin{enumerate}
	    \item {\bf(Latent timeline recovery)} {\it Can we recover a latent disease progression timeline from irregular and heterogeneous ICU EHR trajectories?}

        \item {\bf (Cross-patient registration and subtype discovery)} {\it Can these latent timelines be registered across patients to identify clinically meaningful AKI subtypes?}

        \item {\bf (Downstream utility)} {\it Can the recovered and registered disease timelines improve downstream clinical analyses?}
	\end{enumerate}
    	
	Answering these questions requires an EHR-tailored registration method that respects the sparsity, irregularity, and heterogeneity of ICU data, and scales to large cohorts.
	
	\subsection{Timeline misalignment and statistical formulation}
	
	We consider an EHR dataset comprising $N$ ICU stays (subjects), indexed by $i = 1,\ldots,N$. For each subject $i$, let $\{(t_{ir}, X_i(t_{ir})) : r = 1,\ldots,T_i\}$ denote their longitudinal SCr measurements, where $t_{ir}$ is the time (in days) since ICU admission and $T_i$ is the number of observations. The observation times $t_{ir}$ lie in a common window $[T_{\min}, T_{\max}]$ but are irregular and subject-specific; both the number and timing of measurements vary substantially across patients.
	
	We postulate that patients can be grouped into latent AKI subtypes, with each subtype following a distinct underlying disease progression pattern. Let $\mathcal{P}_1,\ldots,\mathcal{P}_K$ denote $K$ disjoint subtypes, and let $\pi_i \in \{1,\ldots,K\}$ be the subtype label for subject $i$. For a given subtype $k$, we assume there exists an underlying ``intrinsic'' disease timeline such that trajectories within $\mathcal{P}_k$ are approximately aligned after subject-specific time shifts $\delta_i$. Intuitively, $\delta_i$ represents the delay between true disease onset and ICU admission for subject $i$.
	
	Our goal is to estimate both the subtype labels $\{\pi_i\}$ and the time shifts $\{\delta_i\}$ from the observed, irregularly sampled trajectories $\{X_i(t_{ir})\}$, such that the shifted trajectories $\{(t_{ir}+\delta_i,\; X_i(t_{ir}))\}_{r=1}^{T_i}$ are well aligned within each subtype. At the same time, we must be robust to outliers and atypical trajectories, which are common in ICU EHR data \cite{hauskrecht2013outlier}. Formally, we aim to find clustering and shifts that minimize within-subtype dispersion of the aligned trajectories while downweighting or trimming potential outliers.	

	\section{Method for Subtype-Aware Registration of EHR} \label{sec:methods}
	
\begin{figure}[htbp!]
		\centering
		\includegraphics[width=\linewidth]{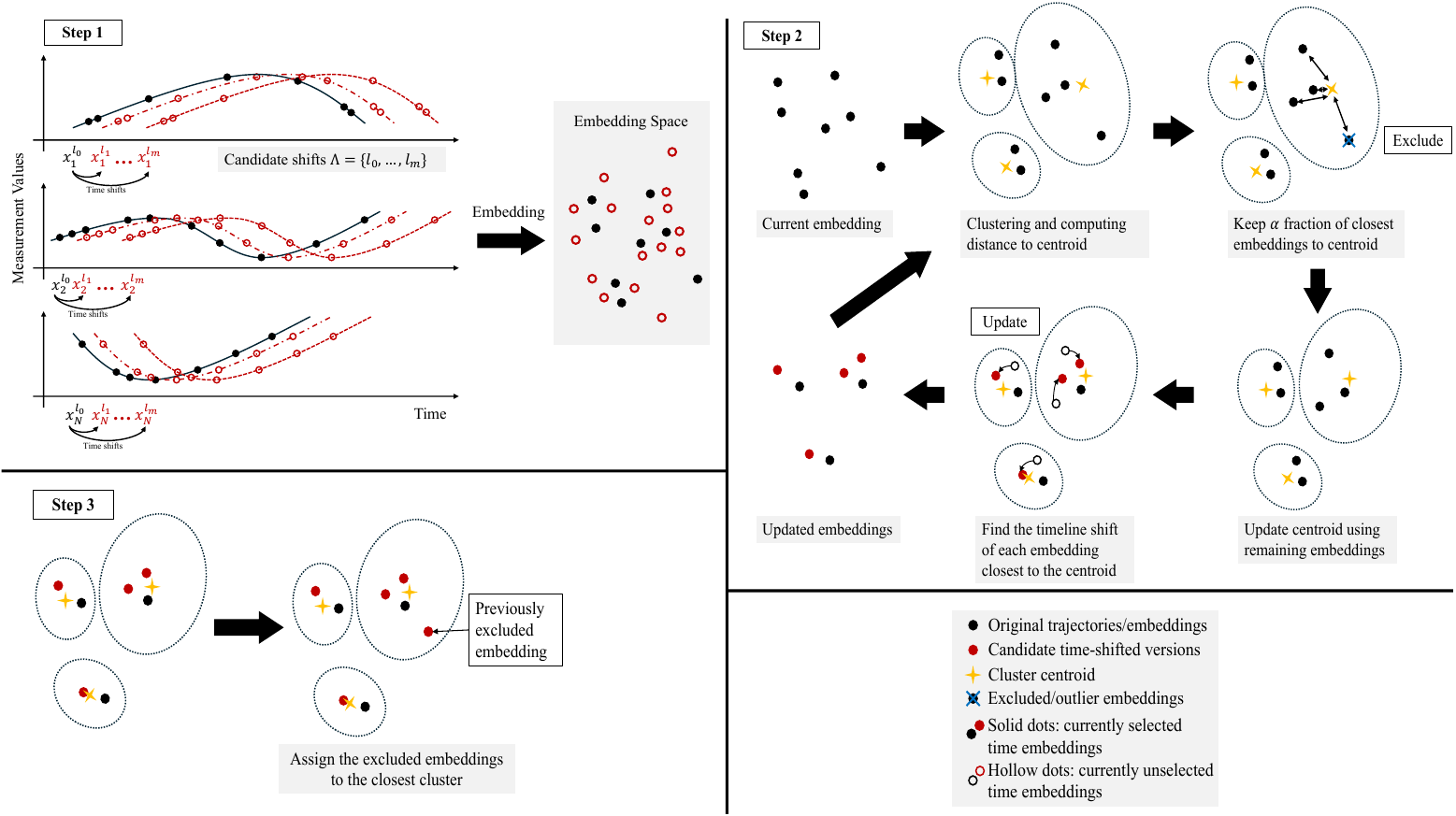}
		\caption{Illustration of subtype-aware registration using longitudinal EHR data.}
		\label{fig:framework}
	\end{figure}

		To address this problem, we develop a scalable, subtype-aware registration algorithm tailored to longitudinal EHR data. The algorithm proceeds in three steps (illustrated in Figure~\ref{fig:framework}):
	
	\begin{enumerate}[leftmargin=*]
		\item \textbf{Spline-based time series embedding.}  
		We first map each irregular trajectory, under a grid of candidate time shifts, to a low-dimensional feature representation using cubic B-spline regression with ridge regularization. This yields a compact representation of each subject–shift combination, greatly reducing the dimensionality and enabling efficient comparison of trajectories with different sampling patterns.
		
		\item \textbf{Iterative clustering and time-shift refinement.}  
		Starting from zero shifts, we repeatedly cluster the current embeddings (using $k$-medoids or $k$-means) and update subject-specific time shifts to bring each trajectory closer to its cluster centroid in the embedding space. At each iteration, we also compute trimmed centroids based on the most central trajectories within each cluster, which improves robustness to outliers and poorly aligned subjects.
		
		\item \textbf{Final assignment of outlying trajectories.}  
		After convergence of the iterative procedure, we reassign previously trimmed or outlying trajectories by jointly optimizing over candidate time shifts and cluster centroids. The final output is a set of estimated shifts $\{\hat{\delta}_i\}$ and subtype labels $\{\hat{\pi}_i\}$ that define aligned AKI progression trajectories for downstream analysis.
	\end{enumerate}

	Next, we describe the proposed registration algorithm in detail. The pseudo-code for each step is provided in Algorithms~\ref{alg:time_series_clustering_1}–\ref{alg:time_series_clustering_3}. Through this paper, we denote $[N]=\{1,\ldots, N\}$ for any integer $N \geq 1$. Let \(|\mathcal{C}|\) be the cardinality of any set \(\mathcal{C}\). The Euclidean norm or the $\ell_2$ norm of a vector $x$ is denoted as \(\left\|\cdot\right\|\), i.e., $\|x\| = \sqrt{\sum_i x_i^2}$. Then, \(\left\|\mathbf{u} - \mathbf{v}\right\|\) represents the Euclidean distance between two vectors \(\mathbf{u}\) and \(\mathbf{v}\).

	\subsection*{Step 1: Time Series Transformation Using Cubic B-spline Regression}
	
	In this step, we apply cubic B-spline regression \citep{wahba1990spline,hastie2009elements} to perform time series transformation. This technique models longitudinal data with flexible and smooth approximations, yielding a low-dimensional representation.
	
	Specifically, we introduce \( q +8\) knots within the overall time domain $[T_{\min}, T_{\max}]$, denoted as $T_{\min} \leq \kappa_1 = \kappa_2 = \kappa_3 = \kappa_4 \leq \cdots \leq \kappa_{q+5} = \kappa_{q+6} = \kappa_{q+7} = \kappa_{q+8} \leq T_{\max}$. Based on these knots, we construct \( p = q + 4 \) basis functions for cubic splines \( B_{j,3}(t) \) recursively as follows.
	
	First, the zeroth-degree basis functions \( B_{j,0}(t) \) are defined as piecewise constants:
	\[
	B_{j,0}(t) = \left\{
	\begin{array}{ll}
		1 & \text{if } t \in [\kappa_j, \kappa_{j+1}), \\
		0 & \text{otherwise}.
	\end{array}
	\right.
	\]
	Then, for \( d = 1, 2, 3 \), the higher-degree basis functions are constructed recursively:
	\[
	B_{j,d}(t) = \frac{t - \kappa_j}{\kappa_{j+d} - \kappa_j} B_{j,d-1}(t) + \frac{\kappa_{j+d+1} - t}{\kappa_{j+d+1} - \kappa_{j+1}} B_{j+1,d-1}(t).
	\]
	For more details about B-splines, readers are referred to \cite{wahba2014spline}. Because the B-spline basis functions sum to one, including an intercept together with all $B_{j,3}(t)$ leads to linear dependence. We therefore include an intercept and drop one basis function $B_{1,3}$ to ensure identifiability. If the boundary knots $\kappa_1, \kappa_{q+8}$ were fixed globally, shifting a patient’s timeline could leave sparse or no observations near the boundaries, so the fitted spline would be weakly constrained by the data there and could become unstable. To mitigate this instability in practice, we can also set the boundary knots to the minimum and maximum observed time points for each shifted trajectory, after applying the candidate time shifts.
	
	We perform B-spline regression using ridge regression (which improves numerical stability) to fit both the original and time-shifted observations for each subject \( i = 1, \ldots, N \). Specifically, we pre-specify a set of potential shift positions \( \Lambda = \{ l_0, l_1, \ldots, l_m \} \), where \( l_0 = 0 \) means no shift. For each \( i = 1, \ldots, N \) and \( l \in \Lambda \), we perform ridge regression \( X_i(t) \sim B_{j,3}(t+l) \) by evaluating the following estimator:
	\begin{equation*}
		\hat{\boldsymbol{\beta}}_{i,l} = \argmin_{\boldsymbol{\beta}_{i,l}} \left\{ \sum_{\substack{r=1 \\ T_{\min} \leq t_{ir} + l \leq T_{\max}}}^{T_i} \left( X_i(t_{ir}) -\beta_{i,l,0} - \sum_{j=2}^{p} \beta_{i,l,j} B_{j,3}(t_{ir}+l) \right)^2 + \lambda \sum_{j=2}^{p} \beta_{i,l,j}^2 \right\},
	\end{equation*}
	where \( \lambda \) is the tuning parameter that controls the penalization level, and \( \hat{\boldsymbol{\beta}}_{i,l} = (\hat{\beta}_{i,l,0},\hat{\beta}_{i,l,2}, \ldots, \hat{\beta}_{i,l,p}) \) is the vector of estimated coefficients. In this way, the underlying structure and dynamics of the longitudinal data are reduced to \( p \) regression coefficients (including an intercept). This cubic B-spline regression yields the following low-dimensional embedding:
	\[
	\{ (t_{ir}, X_i(t_{ir})) \mid r = 1, 2, \ldots, T_i \} \quad \longrightarrow \quad \hat{\boldsymbol{\beta}}_{i,l}.
	\]
	We compute and store these low-dimensional representations for all subjects and potential shifts into a tensor \( \mathbf{\Omega} \in \mathbb{R}^{N \times |\Lambda| \times p} \):
	\[
	\mathbf{\Omega}_{[i, l, :]} = \hat{\boldsymbol{\beta}}_{i,l}, \quad \text{for } i = 1, \ldots, N; \quad l \in \Lambda.
	\]
	
	\subsection*{Step 2: Iterations}
	
	We set the initial time shifts as \( \delta_i^{(0)} = 0 \) for all \( i = 1, \ldots, N \). Next, we run the following steps for \( h = 1, 2, \ldots \):
	\begin{enumerate}[leftmargin=*]
		\item {\bf Clustering with Optimal Number of Clusters.} For each candidate number of clusters \( k \in \{2, \ldots, M\} \), we conduct clustering (e.g., \( k \)-means, \( k \)-medoids) with \( k \) clusters on the low-dimensional representations with time shifts \( \delta_i^{(h-1)} \): $\{ \mathbf{\Omega}_{i, \delta_i^{(h-1)}} \}_{i=1}^N.$
		When \( h = 1 \), the data is the original longitudinal dataset without time shifts.
		
		We determine the number of clusters \( K^{(h)} \) by selecting the \( k \) that maximizes the average silhouette coefficient \citep{rousseeuw1987silhouettes}:
		\begin{equation*} 
			K^{(h)} = \argmax_{k \in \{2, \ldots, M\}} s(k), \quad s(k) = \frac{1}{N} \sum_{i=1}^{N} s(i, k),
		\end{equation*}
		where \( s(i, k) \) is the silhouette coefficient for subject \( i \) when the number of clusters is \( k \), defined as:
\[
s(i, k) = \frac{b(i) - a(i)}{\max\{a(i), b(i)\}}.
\]
Here, \( a(i) \) is the average distance of subject \( i \) to all other subjects within the same cluster and \( b(i) \) is the lowest average distance of subject \( i \) to any other cluster to which \( i \) does not belong, calculated as:
  \[
  a(i) = \frac{1}{|\mathcal{C}_w|-1} \sum_{j \in \mathcal{C}_w, j \neq i} \|\mathbf{\Omega}_{i, \delta_i^{(h-1)}} - \mathbf{\Omega}_{j, \delta_j^{(h-1)}}\|,\quad 
  b(i) = \min_{w' \neq w} \frac{1}{|\mathcal{C}_{w'}|} \sum_{j \in \mathcal{C}_{w'}} \|\mathbf{\Omega}_{i, \delta_i^{(h-1)}} - \mathbf{\Omega}_{j, \delta_j^{(h-1)}}\|,
  \]
  where \( \mathcal{C}_w \) is the cluster to which subject $i$ belongs to. 
		
		This results in \( K^{(h)} \) clusters of the subjects: \( \{ \mathcal{C}_1^{(h)}, \ldots, \mathcal{C}_{K^{(h)}}^{(h)} \} \subseteq [N] \), where \( \mathcal{C}_i^{(h)} \cap \mathcal{C}_j^{(h)} = \emptyset \) for all \( i \neq j \).
		
		\item{\bf Iterative Updates.} For each cluster \( k = 1, \ldots, K^{(h)} \), we perform:
		
		\begin{itemize}[leftmargin=*]
			\item Compute the centroid of cluster \( \mathcal{C}_k^{(h)} \):
			\[
			\bar{\boldsymbol{\beta}}_k^{(h)} = \frac{1}{|\mathcal{C}_k^{(h)}|} \sum_{i \in \mathcal{C}_k^{(h)}} \mathbf{\Omega}_{i, \delta_i^{(h-1)}}.
			\]
			\item Calculate the Euclidean distance between each point in the cluster and the centroid:
			\[
			d_i = \left\| \mathbf{\Omega}_{i, \delta_i^{(h-1)}} - \bar{\boldsymbol{\beta}}_k^{(h)} \right\|, \quad \forall i \in \mathcal{C}_k^{(h)}.
			\]
			
			\item Select the subset \( \mathcal{S}_k^{(h)} \) of points corresponding to the smallest \( \left\lceil \alpha |\mathcal{C}_k^{(h)}| \right\rceil \) distances:
			
			\[
			\mathcal{S}_k^{(h)} = \left\{ i \in \mathcal{C}_k^{(h)} \mid d_i \text{ is among the smallest } \left\lceil \alpha |\mathcal{C}_k^{(h)}| \right\rceil \text{ values} \right\},
			\]
			
			where \( 0 < \alpha \leq 1 \) is a predefined proportion.
			
			\item Recompute the centroid using only the selected points:
			\[
			\bar{\boldsymbol{\gamma}}_k^{(h)} = \frac{1}{|\mathcal{S}_k^{(h)}|} \sum_{i \in \mathcal{S}_k^{(h)}} \mathbf{\Omega}_{i, \delta_i^{(h-1)}}.
			\]
			This step removes potential outliers (\( \mathcal{C}_k^{(h)} \setminus \mathcal{S}_k^{(h)} \)) from the centroid computation, enhancing robustness.
			
			\item Update the time shifts for each subject \( i \in \mathcal{C}_k^{(h)} \) by finding the shift \( \delta_i^{(h)} \in \Lambda \) that minimizes the distance to the updated centroid:
			\[
			\delta_i^{(h)} = \argmin_{\delta \in \Lambda} \left\| \mathbf{\Omega}_{i, \delta} - \bar{\boldsymbol{\gamma}}_k^{(h)} \right\|^2.
			\]
		\end{itemize}
		
		\item {\bf Stopping Criterion.} 

We employ stopping criteria to ensure efficient convergence and to prevent excessive computation. Given an initial clustering \( \{ \mathcal{C}_1^{(1)}, \ldots, \mathcal{C}_{K^{(1)}}^{(1)} \} \), we evaluate the average silhouette coefficient \( s(K^{(1)}) \) using the equation presented earlier in the section. When $s(K^{(1)}) > \tau$ for some predefined threshold $\tau$, we set the maximum number of iterations to be 1 and stop the iterations; otherwise, we run more iterations as follows. 

		At each iteration \( h\geq 2 \), let \( K^{(h)} \) denote the number of clusters selected by maximizing the average silhouette coefficient over a candidate range, and let \( s_{[2]}^{(h)} = \max_{\substack{k \in \{2,\dots,M\}\\k\neq K^{(h)}}} s^{(h)}(k) \) denote the second-highest silhouette score among all \( k \neq K^{(h)} \). The iterative process terminates when the following combined condition is satisfied:
		\[
		\left\{
		K^{(h)} = K^{(h-1)} 
		\;\text{ and }\; 
		s^{(h)}(K^{(h)}) \le s^{(h-1)}(K^{(h-1)}) 
		\;\text{ and }\; 
		s_{[2]}^{(h)} \le s_{[2]}^{(h-1)}
		\right\} \text{ or }
		\left\{ h > Num_{stop} \right\},
		\]
		where \( Num_{stop}  \) is a hard cap on the number of iterations.
		
		This unified condition ensures the iteration breaks if the clustering solution stabilizes, if the initial quality is already sufficient, or if a maximum number of iterations is reached.
		
	\end{enumerate}
	
	\subsection*{Step 3: Finalization}
	Upon termination at iteration \( h = h_{\text{final}} \), we address the data points not selected during the last iteration. For each cluster \( k \), consider the set of unselected points: $\mathcal{C}_k^{(h_{\text{final}})} \setminus \mathcal{S}_k^{(h_{\text{final}})}$,
	where \( \mathcal{S}_k^{(h_{\text{final}})} \) denotes the subset of selected points within cluster \( k \) after the final iteration.
	
	For each unselected point \( j \in \mathcal{C}_k^{(h_{\text{final}})}\setminus\mathcal{S}_k^{(h_{\text{final}})} \), find the optimal shift \( \delta_j^* \) and assign it to the most appropriate cluster center by solving:
	\[
	\delta_j^*,\ k'^* = \argmin_{\delta_j \in \Lambda,\ k' = 1, \ldots, K^{(h_{\text{final}})}} \left\| \mathbf{\Omega}_{j, \delta_j} - \bar{\boldsymbol{\gamma}}_{k'}^{(h_{\text{final}})} \right\|.
	\]
	This assigns $j$ to $\mathcal{S}_{{k'}^{*}}^{(h_{\text final})}$ and shift state that best represents it.
	
	The final output of the algorithm is the number of clusters $\hat{K}^{h_{\rm final}}$, the set of shift states, and clusters for all subjects:
	\[
	\{\delta_i^* : i = 1, 2, \ldots, N \} \quad \text{and} \quad \{\mathcal{S}_k^{(h_{\text{final}})}\}.
	\]
	The overall pseudo code is summarized in Algorithms \ref{alg:time_series_clustering_1}-\ref{alg:time_series_clustering_3}.
	
	\begin{algorithm}[t!] 
		\footnotesize
		\caption{Time Series Clustering with Dynamic Time Shifts: Step 1: Transformation}
		\label{alg:time_series_clustering_1}
		\begin{algorithmic}[1]
			\Require Longitudinal data $\{(t_{ir}, X_i(t_{ir})) \mid i = 1, \ldots, N; \ r = 1, \ldots, T_i\}$,		Potential shift positions $\Lambda = \{l_0, l_1, \ldots, l_m\}$, 
			Maximum number of clusters $M$, predefined threshold $\tau$, 
			Parameters $\alpha$, penalization parameter $\lambda$.
			
			\Ensure Optimal shift states $\{\delta_i^* \mid i = 1, 2, \ldots, N\}$ and clusters $\{\mathcal{C}_k^{(h_{\text{final}})}\}$
			
			\State Introduce  knots $\{\kappa_1, \ldots, \kappa_{q+8}\}$ in $[T_{\min}, T_{\max}]$
			\State Construct cubic B-spline basis functions $B_{j,3}(t)$
			\For{each subject $i = 1$ to $N$}
			\For{each shift $l \in \Lambda$}
			\State Perform ridge regression to obtain coefficients:
			\[
			\hat{\boldsymbol{\beta}}_{i,l} = \argmin_{\boldsymbol{\beta}_{i,l}} \left\{ \sum_{\substack{r=1 \\ T_{\min} \leq t_{ir}+l \leq T_{\max}}}^{T_i} \left( X_i(t_{ir})-\beta_{i,l,0} - \sum_{j=2}^{p} \beta_{i,l,j} B_{j,3}(t_{ir}+l) \right)^2 + \lambda \sum_{j=2}^{p} \beta_{i,l,j}^2 \right\}
			\]
			\State Store coefficients in tensor $\mathbf{\Omega}_{[i, l, :]} = \hat{\boldsymbol{\beta}}_{i,l}$
			\EndFor
			\EndFor
		\end{algorithmic}
	\end{algorithm}

	\begin{algorithm}[t!]
		\footnotesize
		\caption{Step 2: Iterative Clustering and Time Shift Adjustment}
		\label{alg:time_series_clustering_2}
		\begin{algorithmic}[1]
			\State Initialize time shifts $\delta_i^{(0)} = 0$ for all $i$. Set iteration counter $h \gets 1$
			\Repeat
			\For{each $k$ in $\{2, \ldots, M\}$}
			\State Perform clustering with $k$ clusters on $\{\mathbf{\Omega}_{i, \delta_i^{(h-1)}}\}_{i=1}^N$
			\State Compute the average silhouette coefficient $s(k)$
			\EndFor
			\State Select the optimal number of clusters:
			$K^{(h)} = \argmax_{k} s(k)$; obtain clusters $\{\mathcal{C}_1^{(h)}, \ldots, \mathcal{C}_{K^{(h)}}^{(h)}\}$
			\For{each cluster $k = 1$ to $K^{(h)}$}
			\State Compute centroid: $\bar{\boldsymbol{\beta}}_k^{(h)} = \sum_{i \in \mathcal{C}_k^{(h)}} \mathbf{\Omega}_{i, \delta_i^{(h-1)}} / |\mathcal{C}_k^{(h)}| $
			\State Calculate distances $d_i = \left\| \mathbf{\Omega}_{i, \delta_i^{(h-1)}} - \bar{\boldsymbol{\beta}}_k^{(h)} \right\|$ for $i \in \mathcal{C}_k^{(h)}$
			\State Select subset $\mathcal{S}_k^{(h)}$ of points with smallest $\left\lceil \alpha |\mathcal{C}_k^{(h)}| \right\rceil$ distances
			\State Recompute centroid:
			$\bar{\boldsymbol{\gamma}}_k^{(h)} =  \sum_{i \in \mathcal{S}_k^{(h)}} \mathbf{\Omega}_{i, \delta_i^{(h-1)}} / |\mathcal{S}_k^{(h)}|$
			\EndFor
			\For{each subject $i = 1$ to $N$}
			\State Update time shift: $\delta_i^{(h)} = \argmin_{\delta \in \Lambda} \left\| \mathbf{\Omega}_{i, \delta} - \bar{\boldsymbol{\gamma}}_k^{(h)} \right\|^2$,  where $k$ is the cluster containing subject $i$
			\EndFor
			
			\State \textbf{Check Stopping Criteria:}
\If {$h=1$ and $s(K^{(1)}) > \tau$}
\State Terminate
\EndIf
\State $K^{(h)} \leftarrow \arg\max_k s^{(h)}(k)$; $s^{(h)} \leftarrow s^{(h)}(K^{(h)})$; $s_{[2]}^{(h)} \leftarrow \max_{\substack{k \in \{2,\dots,M\} \\ k \neq K^{(h)}}} s^{(h)}(k)$
\If{\{$K^{(h)} = K^{(h-1)}$,  $s^{(h)}(K^{(h)}) \le s^{(h-1)}(K^{(h-1)})$ and $s_{[2]}^{(h)} \le s_{[2]}^{(h-1)}$\} or \{$h > Num_{stop}$\}}
\State Terminate
\EndIf
	\Until{Termination condition is met}
		\end{algorithmic}
	\end{algorithm}

	\begin{algorithm}[t!]
		\footnotesize
		\caption{Step 3: Assign Unselected Points and Finalize}
		\label{alg:time_series_clustering_3}
		\begin{algorithmic}[1]
			\For{each cluster $k$}
			\For{each unselected point $j \in \mathcal{C}_k^{(h_{\text{final}})} \setminus \mathcal{S}_k^{(h_{\text{final}})}$}
			\State Assign optimal shift and cluster by evaluating
			\[
			\delta_j^*,\ k'^* = \argmin_{\delta_j \in \Lambda,\ k' = 1, \ldots, K^{(h_{\text{final}})}} \left\| \mathbf{\Omega}_{j, \delta_j} - \bar{\boldsymbol{\gamma}}_{k'}^{(h_{\text{final}})} \right\|.
			\]
			Then reassign $j$ to cluster $\mathcal{S}_{{k'}^{*}}^{(h_{\text{final}})}$.
			\EndFor
			\EndFor
			\State \textbf{Output} final shift amounts $\{\delta_i^* \mid i = 1, 2, \ldots, N\}$ and clusters $\{\mathcal{S}_k^{(h_{\text{final}})}\}$
		\end{algorithmic}
	\end{algorithm}

	\section{Comprehensive Analyses of Longitudinal EHR Data in Acute Kidney Injury Patients}\label{sec:real-data}
	
	In this section, we apply the proposed registration method to longitudinal SCr trajectories from the MIMIC-IV ICU AKI cohort described in Section~\ref{sec:data_problem}.
	
	\subsection{Experimental Setup} \label{sec:setup}

	\paragraph{Dataset and Cohort Selection}
	We use the MIMIC-IV AKI ICU cohort defined in Section~\ref{sec:data_problem}, consisting of ICU stays with longitudinal serum creatinine (SCr) trajectories over the first 21 days after ICU admission.

	\paragraph{Data Preprocessing and Model Selection}
	
	We perform unsupervised learning on the data. Specifically, we transform the original longitudinal data as detailed in Section~\ref{sec:methods}, and used the estimated coefficients of the spline basis functions to represent each ICU stay. We applied \( k \)-means clustering \cite{macqueen1967some} and \( k \)-medoids clustering \cite{rdusseeun1987clustering} on these representations. The quality of the resulting clusters was evaluated by computing the average silhouette coefficient \cite{rousseeuw1987silhouettes}.
	
	For clinical evaluation, we labeled each ICU stay based on the AKI severity stages defined below \cite{Khwaja2012KDIGOCP,Zhu2024FunctionalPS}:
	\begin{enumerate}[leftmargin=*]
		\item \textbf{Stage 1}: Serum creatinine (SCr) level rises to \( 1.5 \times \) to \( 1.9 \times \) the baseline value within seven days since time 0.
		\item \textbf{Stage 2}: SCr level rises to \( 2.0 \times \) to \( 2.9 \times \) the baseline value within seven days  since time 0.
		\item \textbf{Stage 3}: SCr level rises to \( 3 \times \) or greater than the baseline value within seven days, or the maximum SCr level over two days is \( \geq 4.0 \; \text{mg/dL} \). We further divide Stage 3 into two subtypes \cite{Zhu2024FunctionalPS}:
		\begin{enumerate}[leftmargin=*]
			\item \textbf{Subtype 1}: SCr level \( \geq 4.0 \; \text{mg/dL} \) within 48 hours  since time 0.
			\item \textbf{Subtype 2}: SCr level rises to \( 3 \times \) or greater than the baseline value within seven days since time 0 and does not meet the criteria for Subtype 1.
		\end{enumerate}
	\end{enumerate}
	The baseline SCr value was computed using the Modification of Diet in Renal Disease (MDRD) equation \cite{Levey1999AMA}.
	
	We also perform supervised learning on the data. We specifically organize the original longitudinal EHR data into an \( N \times T \) array, where \( N \) is the number of ICU stays and \( T \) represents the number of days since ICU admission. We performed data normalization and missing data imputation following the methods described in \cite{Jiang2023SoftPF}. We computed summary statistics (maximum, minimum, mean, standard deviation, and skewness) over 3-day and 4-day windows and performed severe AKI prediction using logistic regression, based on the methods described in \cite{Lipton2016LearningTD}. We defined {\it severe AKI} as observing a \( 3 \times \) increase in creatinine level compared to the baseline value or observing any creatinine level \( \geq 4.0 \; \text{mg/dL} \) during the whole observed time interval of 21 days \cite{Khwaja2012KDIGOCP}. For the baseline method in \cite{jiang2023timeline}, we set the potential time shifts to within seven days and the maximum iteration number to 10.

	\subsection{Unsupervised Learning -- Phenotyping}\label{sec:uspl}
	We first evaluate the impact of timeline registration on clustering AKI ICU stays using \( k \)-medoids and \( k \)-means. The average silhouette coefficients for different numbers of clusters \( K \in [2, 8] \) are presented in Table~\ref{tab:unsup-data}. Notably, when \( K = 2 \) and \( K = 3 \), the clustering results yield significantly higher average silhouette coefficients, indicating more distinct clustering results. Additionally, we observe that both \( k \)-medoids and \( k \)-means show substantial increases in the average silhouette coefficients after applying the proposed timeline registration method. Given the heterogeneity of AKI, we expect the ICU stays to cluster according to different subtypes of the condition. The numerical results demonstrate the effectiveness of our timeline registration method in aligning the timelines of patients within the same subtype. Furthermore, our proposed registration method clearly outperforms the baseline registration method, achieving more distinct subtype separation for most choices of \( K \).
	\begin{table}[ht!]
  \centering
  \resizebox{\textwidth}{!}{
  \begin{tabular}{@{}lllllll@{}}
    \toprule
    & \multicolumn{3}{l}{$k$-medoids} & \multicolumn{3}{l}{$k$-means} \\ 
    \midrule
    & Observed Data 
      & Registered Data 
      & \makecell{Registered Data\\(baseline)} 
      & Observed Data 
      & Registered Data 
      & \makecell{Registered Data\\(baseline)} \\ 
    \midrule
    $K$=2 & 0.431 & 0.481 & 0.266 & 0.584 & 0.621 & 0.518 \\
    $K$=3 & 0.226 & 0.261 & 0.326 & 0.523 & 0.557 & 0.497 \\ 
    $K$=4 & 0.249 & 0.276 & 0.276 & 0.325 & 0.328 & 0.479 \\ 
    $K$=5 & 0.208 & 0.278 & 0.086 & 0.350 & 0.380 & 0.307 \\ 
    $K$=6 & 0.208 & 0.282 & 0.075 & 0.352 & 0.394 & 0.330 \\
    $K$=7 & 0.111 & 0.164 & 0.075 & 0.346 & 0.239 & 0.328 \\
    $K$=8 & 0.108 & 0.156 & 0.025 & 0.284 & 0.220 & 0.296 \\ 
    \bottomrule
  \end{tabular} }
  \caption{Comparison of average silhouette coefficients for original and registered data after applying the proposed timeline registration method, using \(k\)-medoids and \(k\)-means clustering.}
  \label{tab:unsup-data}
\end{table}

	In computational phenotyping research, the number of subtypes \( K \) is often unknown and must be determined. Therefore, we consider a scenario in which the number of clusters falls within a pre-specified range, and we apply the proposed timeline registration method to determine this number in a data-driven manner. In Table~\ref{tab:unsup-spec}, we present results for two scenarios: one requiring a smaller number of clusters and the other requiring a larger number, as potentially suggested by auxiliary information. In both cases, we observe that the maximum average silhouette coefficient increased after timeline registration, indicating that the timeline registration method provides consistent improvement in downstream clustering performance.
	\begin{table}[!ht]
		\centering
		{
			\begin{tabular}{@{}lllll@{}}
				\toprule
				& \multicolumn{2}{l}{$k$-medoids} & \multicolumn{2}{l}{$k$-means} \\ \midrule
				& Observed Data & Registered Data & Observed Data & Registered Data \\
				\( K \in [2, 4] \) & 0.431 & 0.482 & 0.584 & 0.621 \\
				\( K \in [5, 8] \) & 0.208 & 0.284 & 0.352 & 0.432 \\ \bottomrule
			\end{tabular}
		}
		\caption{Comparison of the maximum average silhouette coefficients obtained within the cluster ranges \(K\in[2,4]\) and \(K\in[5,8]\) for the original data and the registered data after applying the proposed timeline registration method, using \(k\)-medoids and \(k\)-means clustering.}
		\label{tab:unsup-spec}
	\end{table}

    \subsubsection{Phenotype Interpretation and Evaluation}
Next, we further explored and evaluated the phenotyping results before and after temporal registration. From a precision-medicine perspective, clinically meaningful phenotypes are not only associated with distinct biology and outcomes, but also need to be identifiable early enough in the disease course to inform prevention, timely interventions, and resource allocation \cite{matthay2020phenotypes, wick2021promises, mcdonald2019treatable}. For clarity, we denote phenotypes obtained from the raw (unregistered) trajectories by $S_k^{\mathrm{raw}}$ and those obtained from the time-registered trajectories by $S_k^{\mathrm{rec}}$.

\begin{figure}[!hbt]
    \centering
    \includegraphics[width=.99\textwidth]{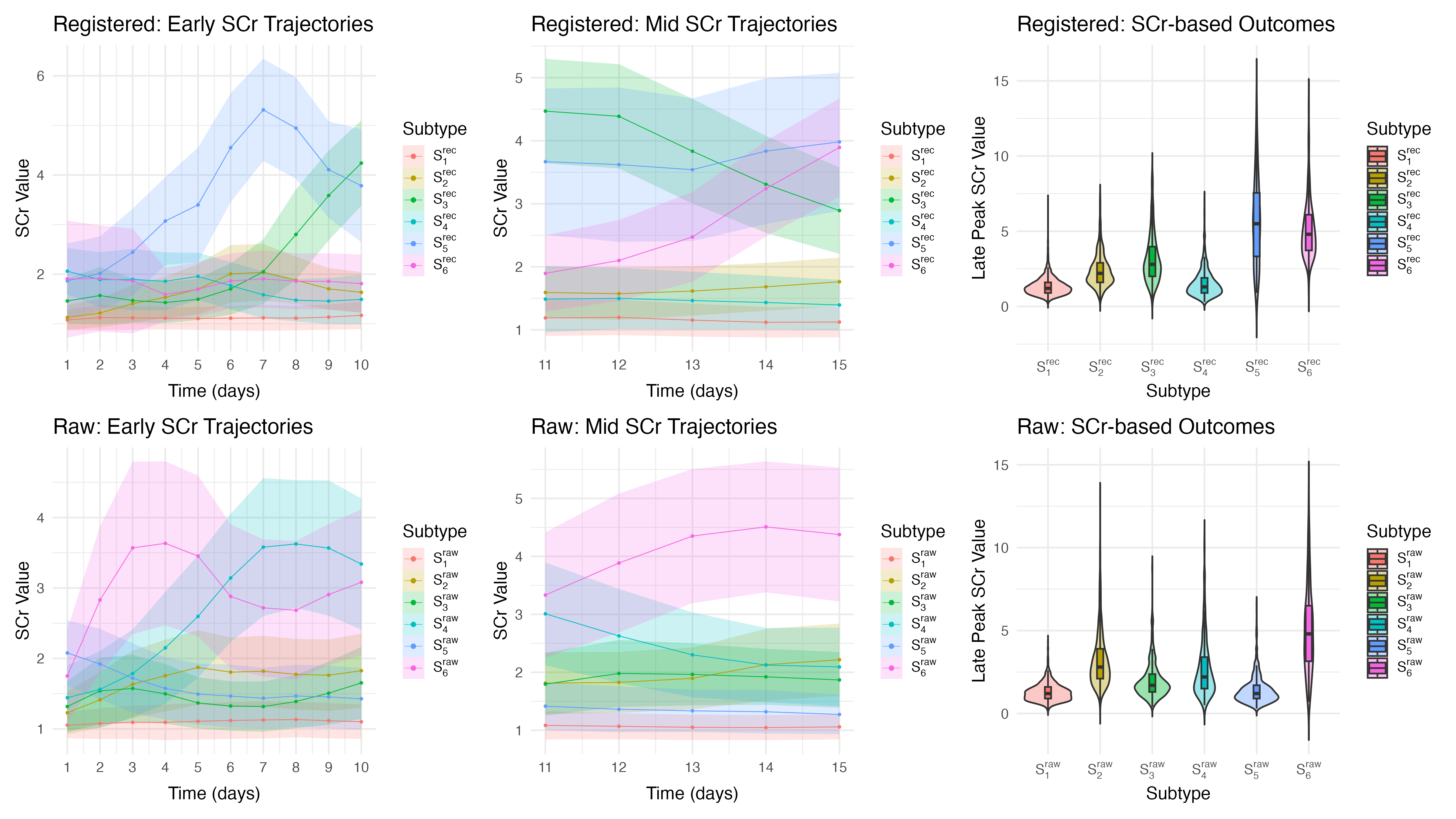}
    \caption[Early, mid and late SCr trajectories and SCr-based outcome by subtype.]%
    {Early and mid SCr trajectories and SCr-based outcome by subtype. For the early and mid periods (definitions given in the Methods), the curves display the mean SCr over time and the shaded bands represent mean $\pm 0.5$ SD within each subtype. The right-hand panels summarise the SCr-based outcome, defined as each patient's peak SCr, i.e., the maximum SCr value observed in the late period.}
    \label{fig:SCr_registered_raw}
\end{figure}

Based on the unsupervised clustering results described above, we focused on the clustering solution with $K=6$, which achieved the largest average silhouette coefficient among the more fine-grained clusterings considered (i.e., for $K \ge 3$). In this analysis, we divided the 21-day observation window into an early period (days 1--10), a mid period (days 11--15), and a late period (days 16--21), and defined a SCr-based outcome as the maximum creatinine value observed in the late period. We then evaluated how reliably phenotypes could be identified in the early and intermediate windows. As shown in Figure \ref{fig:SCr_registered_raw}, for the raw data, the two phenotypes that can be identified in the early and mid windows are $S_4^{\mathrm{raw}}$ and $S_6^{\mathrm{raw}}$, and only $S_6^{\mathrm{raw}}$ remains clearly associated with a distinct SCr-based outcome. In contrast, for the registered data the three phenotypes that are most easily identifiable in the early and mid windows are $S_3^{\mathrm{rec}}$, $S_5^{\mathrm{rec}}$, and $S_6^{\mathrm{rec}}$, and all three exhibit well-separated distributions of this SCr-based outcome. 

These findings suggest that phenotyping based on the registered data yields phenotypes that are more clinically meaningful and potentially more actionable: once a longitudinal phenotype can be recognized early, we gain phenotype-specific information about the subsequent course, which in turn can guide targeted treatment and nursing strategies and facilitate the implementation of personalized care.

To further evaluate these phenotypes, we used diagnosis text variables to characterize the three time-registered phenotypes. We first excluded diagnosis entries that explicitly referred to ``acute kidney injury'' or ``acute kidney failure'', so that AKI-defining terms themselves were not used in the downstream characterization. Among the remaining diagnoses, we then performed a focused semantic harmonization for a set of predefined, clinically important diagnosis terms by grouping them into several categories, including hyperkalemia, hyponatremia and/or hyposmolality, acute respiratory failure, the sepsis spectrum, essential (primary) hypertension, hyperlipidemia, atrial fibrillation, ascites, urinary tract infection, and chronic kidney disease, mapping all synonymous textual descriptions of each of these conditions to the corresponding category. Using these harmonized categories, we summarized the prevalence of key diagnoses (proportion of patients with each diagnosis) across $S_3^{\mathrm{rec}}$, $S_5^{\mathrm{rec}}$, and $S_6^{\mathrm{rec}}$ (Figure \ref{fig:diag_rec_S3S5S6}), which provided a concise overview of their diagnostic profiles. This led to the following clinical interpretation of these three AKI phenotypes:

\begin{figure}[!hbt]
    \centering
    \includegraphics[width=.99\textwidth]{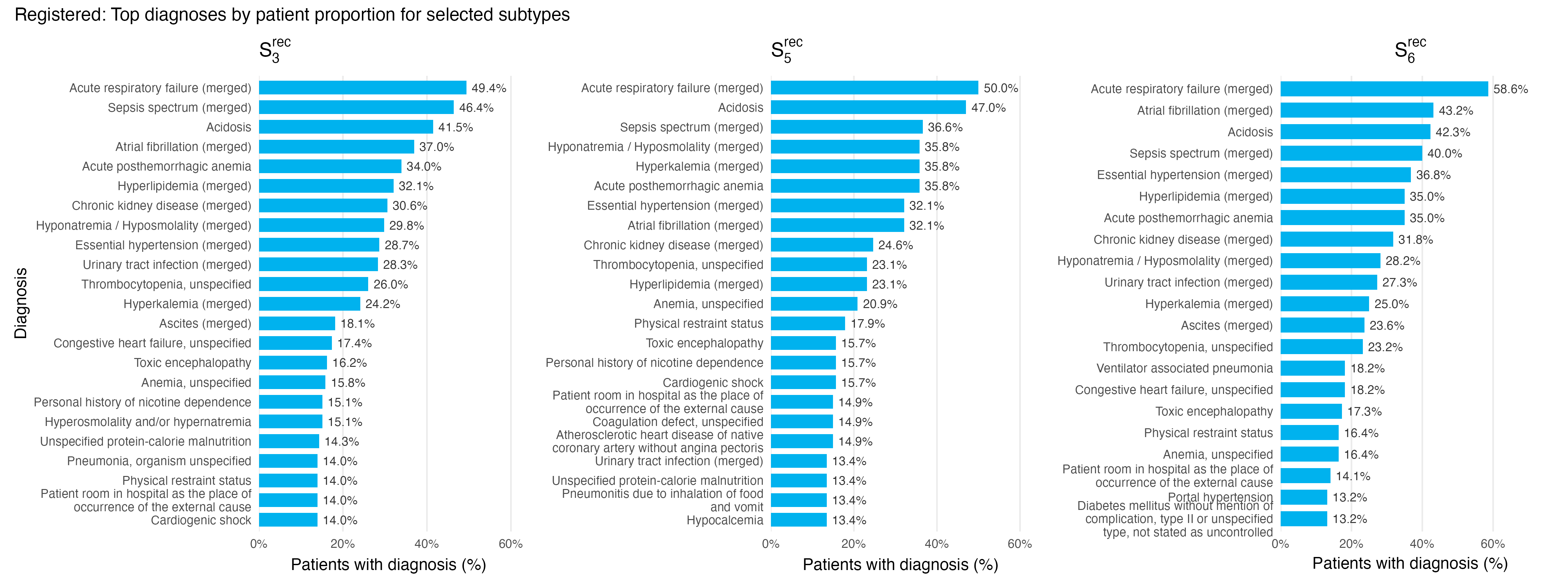}
    \caption[Top diagnoses for registered subtypes.]%
    {Top 20 diagnoses by patient proportion for the registered subtypes
    $S_3^{\mathrm{rec}}$, $S_5^{\mathrm{rec}}$, and $S_6^{\mathrm{rec}}$.}
    \label{fig:diag_rec_S3S5S6}
\end{figure}

Phenotype $S_3^{\mathrm{rec}}$ (a sepsis-dominant AKI phenotype) is characterized by a high burden of sepsis-spectrum diagnoses together with acute respiratory failure, metabolic acidosis, and moderate electrolyte disturbances (hyponatremia/hyposmolality). Its clinical profile is broadly similar to the clinical picture of sepsis-associated AKI described in the recent ADQI consensus report by Zarbock et al. \cite{zarbock2023sepsis}.

Phenotype $S_5^{\mathrm{rec}}$ (a metabolic-electrolyte derangement AKI phenotype) is characterized by a high prevalence of acute respiratory failure, acidosis, hyponatremia/hyposmolality, hyperkalemia, and acute post-hemorrhagic anemia. This clinical profile shares key features with critically ill patients in whom AKI is accompanied by severe metabolic and electrolyte complications, regarded as classic indications for urgent kidney replacement therapy \citep{rahman2012acute}.

Phenotype $S_6^{\mathrm{rec}}$ (a cardiopulmonary failure with multi-organ dysfunction AKI phenotype) shows frequent acute respiratory failure and relatively high prevalence of atrial fibrillation, sepsis-spectrum diagnoses, metabolic acidosis, chronic kidney disease, and other organ dysfunctions. Its clinical profile is close to AKI occurring with acute cardiopulmonary failure (for example, acute decompensated heart failure as a typical cardiorenal syndrome type~1 scenario) and with acute respiratory failure/ARDS, where lung-kidney cross-talk has been linked to organ dysfunction and adverse outcomes \citep{rangaswami2019cardiorenal, HusainSyed2016LungKidney}.

Taken together, these similarities indicate that temporal registration substantially improves phenotyping: after registration, the clustering recovers clinically meaningful AKI patterns that can be used as interpretable building blocks in subsequent analyses.
    	
	\subsection{Clinical Evaluation} \label{sec:clinic}
    
	Next, we incorporate clinical information, specifically the AKI severity stages, into the evaluation of timeline registration performance. Recall that the definition of AKI severity stages, as outlined in Section~\ref{sec:setup}, relies on the starting time point. This implies that the AKI severity stage of an individual subject may change after applying timeline registration. Therefore, for both the observed data (without timeline registration) and the data after applying the proposed timeline registration procedure, we label each subject as Stage~1~(S1), Stage~2~(S2), Stage~3 Subtype~1~(S3-1), or Stage~3 Subtype~2~(S3-2). We then group both the observed and registered datasets into five different configurations, as described in the first column of Table~\ref{tab:clinical-new}, and evaluate the average silhouette coefficient for each configuration. The results for these five different AKI sub-category divisions are presented in Table~\ref{tab:clinical-new}. Consistently, the data processed through timeline registration demonstrate an increased average silhouette coefficient, indicating improved data aggregation within each AKI sub-category.  The baseline registration method also shows good performance in the first four configurations. However, it underperforms compared to the observed data in the fifth configuration. In contrast, the proposed registration method consistently outperforms the observed data across all configurations, demonstrating the robustness of our approach.
	\begin{table}[ht!]
	\centering
    \resizebox{\textwidth}{!}{
	\begin{tabular}{lcccc}
		\hline
		& Observed Data & Registered Data (baseline) & Registered Data ($k$-means) & Registered Data ($k$-medoids) \\ \hline
		4-Groups (S1, S2, S3-1, S3-2) & -0.049 & -0.038 & -0.039 & -0.041 \\
		3-Groups (S1, S2, (S3-1, S3-2)) & -0.057 & -0.040 & -0.046 & -0.048 \\
		3-Groups ((S1, S2), S3-1, S3-2) & 0.286  & 0.335 & 0.326 & 0.322 \\
		2-Groups ((S1, S2), (S3-1, S3-2)) & 0.321 & 0.341 & 0.342 & 0.338 \\
		2-Groups ((S1, S2, S3-1), S3-2) & 0.527 & 0.503 & 0.625 & 0.663 \\
		\hline
	\end{tabular}
    }
	\caption{Comparison of observed data and registered data in the clinical setting using the new criteria for defining creatinine baseline level. For each AKI sub-category configuration, the reported values are the average silhouette coefficients. Larger values indicate better aggregation and clearer separation of subjects within the specified AKI sub-categories.}
	\label{tab:clinical-new}
\end{table}

	\subsection{Supervised Learning -- Severe AKI Prediction}\label{sec:spl}
	
	To validate the effectiveness of timeline registration in enhancing downstream clinical analysis, we perform severe AKI\footnote{See definition of severe AKI in Data Preprocessing and Model Section in Section \ref{sec:setup}.} prediction and compare the performance using observed data and the recovered data after registration. The experiment was conducted 500 times and we report the mean value with its 95\% confidence interval, presented in Table \ref{tab:sup:3d} and Table \ref{tab:sup:4d}.
	\begin{table}[hbt!]
		\centering
		\resizebox{\textwidth}{!}{
		\begin{tabular}{@{}l l l l l@{}}
			\toprule
			& Observed Data & Registered Data (baseline) & Registered Data ($k$-medoids) & Registered Data ($k$-means) \\ \midrule
			AUROC & 0.8047 [0.7801,0.8275] & 0.7441 [0.7202,0.7687] & 0.8559 [0.8357,0.8734] & 0.8467 [0.8243,0.8679] \\
            Precision & 0.8016 [0.7765,0.827] & 0.8326 [0.7621,0.8637] & 0.8235 [0.801,0.8454] & 0.8181 [0.7942,0.841] \\
            Recall & 0.7139 [0.6944,0.7366] & 0.5114 [0.5056,0.5171] & 0.7546 [0.7328,0.7752] & 0.7454 [0.7241,0.7654] \\
			Accuracy & 0.8167 [0.8031,0.8312] & 0.7268 [0.7234,0.7303] & 0.8385 [0.8247,0.8514] & 0.8333 [0.8193,0.8473] \\
			AUPRC & 0.8022 [0.7803,0.8227] & 0.7393 [0.7115,0.7635] & 0.8446 [0.824,0.8618] & 0.8355 [0.8139,0.8566] \\ \bottomrule
		\end{tabular}
        }
		\caption{Comparisons of severe AKI prediction using 3-day observed data and recovered data after registration. Metrics are presented using mean values with 95\% confidence intervals.}
		\label{tab:sup:3d}
	\end{table}
	
	\begin{table}[hbt!]
		\centering
		\resizebox{\textwidth}{!}{
			\begin{tabular}{@{}l l l l l@{}}
				\toprule
				& Observed Data & Registered Data (baseline) & Registered Data ($k$-medoids) & Registered Data ($k$-means) \\ \midrule
				AUROC & 0.8274 [0.8033,0.8497] & 0.7752 [0.7493,0.7991] & 0.8788 [0.8605,0.895] & 0.8687 [0.8475,0.8875] \\
				Precision & 0.8143 [0.7896,0.84] & 0.7357 [0.7123,0.7607] & 0.8351 [0.8134,0.855] & 0.8353 [0.8146,0.8578] \\
				Recall & 0.733 [0.7121,0.7531] & 0.7127 [0.6897,0.7365] & 0.7721 [0.7515,0.7929] & 0.7694 [0.7495,0.7896] \\
				Accuracy & 0.8276 [0.8138,0.8412] & 0.7862 [0.7682,0.8032] & 0.8488 [0.8357,0.8617] & 0.848 [0.8343,0.861] \\
				AUPRC & 0.8257 [0.8043,0.8472] & 0.7597 [0.7329,0.7849] & 0.8689 [0.8509,0.8846] & 0.8595 [0.8394,0.8775] \\ \bottomrule
			\end{tabular}
		}
		\caption{Comparisons of severe AKI prediction using 4-day observed data and recovered data after registration. Metrics are presented using mean values with 95\% confidence intervals.}
		\label{tab:sup:4d}
	\end{table}

    From both tables, we observe that using data after registration results in significant increases in the AUROC (area under the receiver operating characteristic curve) and AUPRC (area under the precision-recall curve) values. Accuracy values also slightly increase. This comparison indicates that timeline registration is influential in enhancing downstream clinical analysis and can be considered an essential preprocessing step. We also observe that the baseline EHR registration method yields results worse than using data without registration from both tables. This can be attributed to the reason that the baseline method relies on the assumption that all patients in the cohort share the same disease progression timeline. Since AKI patients may undergo distinct disease progression patterns, enforcing patient data to follow a shared trend may result in inaccurate timeline alignment between patients and thus degrading the prediction performance.

   \subsubsection{Subgroup Analysis}
   
   We conducted subgroup analysis on groups by age levels \cite{Xu2021SystematicRA} (group of patients that are below 60 years of old and group of patients over 60 years of old) and gender. We present the performance of the prediction model on observed data and data recovered by our proposed method in each subgroup shown in Table \ref{tab:subgr:gender} and Table \ref{tab:subgr:age}. The experiments were conducted 500 times using 4-day data with the mean values and 95\% confidence intervals reported.

   We observe from the tables that the prediction model yielded higher performance using registered data obtained from the proposed method than using the original data in both gender groups. Likewise, we note a higher performance of the model with the use of proposed data registration in both age groups. Additionally, we show the prediction model yielded comparable performance in all subgroups. 
   
   Overall, early warning of AKI across diverse clinical settings is clinically important \cite{song2020cross, zhang2025development}, and our results suggest that the proposed temporal registration method can improve the performance of such supervised prediction models.

    \begin{table}[hbt!]
    \centering
	\resizebox{\textwidth}{!}{
	\begin{tabular}{lllllll}
	\toprule
    & & AUROC & Precision & Recall & Accuracy & AUPRC \\ \midrule
    \multirow{4}{*}{Male} & Observed Data & 0.8357 [0.8088,0.8625] & 0.7974 [0.7659,0.8271] & 0.7459 [0.7174,0.7733] & 0.8318 [0.8124,0.8511] & 0.833 [0.8055,0.8583] \\
    & Registered Data (baseline) & 0.8018 [0.7692,0.8297] & 0.7573 [0.7297,0.7864] & 0.7507 [0.7235,0.7776] & 0.8104 [0.7893,0.8305] & 0.7791 [0.7421,0.8108] \\
    & Registered Data ($k$-medoids) & 0.8983 [0.8755,0.918] & 0.839 [0.8109,0.8638] & 0.8083 [0.7808,0.8319] & 0.867 [0.8482,0.8855] & 0.8865 [0.8642,0.9066] \\
    & Registered Data ($k$-means) & 0.8842 [0.8635,0.9057] & 0.8295 [0.8028,0.857] & 0.7981 [0.7705,0.8217] & 0.8599 [0.8415,0.8797] & 0.8739 [0.8535,0.8941] \\
    \multirow{4}{*}{Female} & Observed Data & 0.8288 [0.7935,0.8631] & 0.8564 [0.8259,0.8836] & 0.7178 [0.6865,0.7496] & 0.821 [0.7975,0.8463] & 0.8327 [0.7992,0.8639] \\
    & Registered Data (baseline) & 0.7448 [0.7056,0.7862] & 0.7007 [0.6649,0.7415] & 0.6594 [0.6268,0.6951] & 0.7472 [0.7167,0.774] & 0.74 [0.7043,0.7801] \\
    & Registered Data ($k$-medoids) & 0.8641 [0.8332,0.8906] & 0.8386 [0.8074,0.8679] & 0.723 [0.6917,0.7531] & 0.8196 [0.796,0.8428] & 0.858 [0.8271,0.8835] \\
    & Registered Data ($k$-means) & 0.8594 [0.8249,0.8921] & 0.8599 [0.8274,0.8868] & 0.7317 [0.7006,0.7638] & 0.8288 [0.8067,0.8536] & 0.8558 [0.8244,0.8867] \\ \cmidrule(l){1-7}
	\end{tabular}
	}
	\caption{Subgroup analysis on gender. Metrics are presented using mean values with 95\% confidence intervals.}
	\label{tab:subgr:gender}
	\end{table}

    \begin{table}[hbt!]
    \centering
	\resizebox{\textwidth}{!}{
	\begin{tabular}{lllllll}
	\toprule
    & & AUROC & Precision & Recall & Accuracy & AUPRC \\ \midrule
    \multirow{4}{*}{Age $< 60$} & Observed Data & 0.8478 [0.8139,0.8838] & 0.818 [0.7766,0.8576] & 0.7662 [0.7332,0.8038] & 0.8483 [0.8241,0.875] & 0.8479 [0.8159,0.8795] \\
    & Registered Data (baseline) & 0.8099 [0.7724,0.8444] & 0.7658 [0.7305,0.8012] & 0.748 [0.7103,0.7859] & 0.8185 [0.7924,0.8439] & 0.7871 [0.7428,0.825] \\
    & Registered Data ($k$-medoids) & 0.9007 [0.8731,0.9265] & 0.8359 [0.8001,0.8687] & 0.7955 [0.761,0.832] & 0.8641 [0.8402,0.8868] & 0.8888 [0.8623,0.9147] \\
    & Registered Data ($k$-means) & 0.8945 [0.866,0.9219] & 0.8429 [0.8102,0.8747] & 0.8123 [0.7795,0.845] & 0.8718 [0.848,0.896] & 0.8875 [0.8604,0.9145] \\
    \multirow{4}{*}{Age $\ge 60$} & Observed Data & 0.8163 [0.7883,0.8432] & 0.8137 [0.7866,0.8403] & 0.7159 [0.6898,0.7433] & 0.8153 [0.7974,0.8341] & 0.8139 [0.7857,0.8391] \\
    & Registered Data (baseline) & 0.7551 [0.7241,0.7841] & 0.7178 [0.6869,0.7484] & 0.6932 [0.6657,0.7221] & 0.7669 [0.7433,0.789] & 0.7446 [0.7117,0.7754] \\
    & Registered Data ($k$-medoids) & 0.8664 [0.8424,0.8871] & 0.8354 [0.8079,0.8613] & 0.76 [0.7337,0.7861] & 0.8397 [0.8213,0.8593] & 0.8579 [0.8348,0.8793] \\
    & Registered Data ($k$-means) & 0.8556 [0.8262,0.8808] & 0.8327 [0.8058,0.859] & 0.7472 [0.7201,0.7739] & 0.8337 [0.8153,0.8529] & 0.8451 [0.8177,0.8692] \\ \cmidrule(l){1-7}
	\end{tabular}
	}
	\caption{Subgroup analysis on age levels. Metrics are presented using mean values with 95\% confidence intervals.}
	\label{tab:subgr:age}
	\end{table}
   
	\section{Discussions and Conclusions} \label{sec:discussions}
	In this paper, we demonstrate the necessity of performing timeline registration for longitudinal EHR data, aligning patient records to reflect the intrinsic disease progression trends. To evaluate the feasibility of the proposed registration algorithm for recovering ground truth longitudinal data, we performed simulations with eight different scenarios. The results indicate that the proposed method can effectively recover observed data to ground truth with low time complexity. 
    To demonstrate the impact of registration on downstream clinical data analyses, we perform phenotyping, clinical evaluation and severe AKI prediction and compare their results using data with or without timeline registration. The results suggest that registration can effectively improve the performance of downstream clinical data analysis. We also conclude that the registration algorithm is of high stability, reflected by the minimal 95\% confidence intervals given 500 times repetition of experiments. Moreover, the proposed framework is flexible and readily extensible. Both the dimension-reduction and clustering steps can be substituted with alternative methods according to users' preferences and practical needs, provided that the chosen dimension-reduction method preserves separability among distinct trajectory patterns and the clustering method is capable of yielding a meaningful grouping structure. This flexibility is empirically illustrated in Supplementary Section~A through ablation studies that examine representative alternatives for these components. On the other hand, our method also has some limitations. For example, in diseases with rare trajectory subtypes, the accuracy of alignment for these rare subtypes may be less accurate due to the potential overshadowing effect from other more dominant subtypes. Registration on continuous space and evaluation of registration on other critical medical events will be investigated in our future work. EHR timeline registration is an emerging field and our work provides insight for improving clinical analysis using EHR data.

	\bibliographystyle{unsrt}
	\bibliography{ref}

\begin{thebibliography}{10}

\bibitem{Shickel2017DeepEA}
Benjamin Shickel, Patrick~James Tighe, Azra Bihorac, and Parisa Rashidi.
\newblock {Deep EHR: A Survey of Recent Advances in Deep Learning Techniques
  for Electronic Health Record (EHR) Analysis}.
\newblock {\em IEEE Journal of Biomedical and Health Informatics},
  22:1589--1604, 2017.

\bibitem{Yadav2017MiningEH}
Pranjul Yadav, Michael~S. Steinbach, Vipin Kumar, and Gy{\"o}rgy~J. Simon.
\newblock {Mining Electronic Health Records: A Survey}.
\newblock {\em ACM Computing Surveys (CSUR)}, 50(6):85, 2018.

\bibitem{Shillan2019UseOM}
Duncan Shillan, Jonathan A.~C. Sterne, Alan~R. Champneys, and Ben Gibbison.
\newblock Use of machine learning to analyse routinely collected intensive care
  unit data: a systematic review.
\newblock {\em Critical Care}, 23:284, 2019.

\bibitem{Torres2021Pneumonia}
Antoni Torres, Catia Cill{\'o}niz, Michael~S. Niederman, Rosario Men{\'e}ndez,
  James~D. Chalmers, Richard~G. Wunderink, and Tom van~der Poll.
\newblock Pneumonia.
\newblock {\em Nature Reviews Disease Primers}, 7:25, 2021.

\bibitem{Irlmeier2022Cox}
Rebecca Irlmeier, Jacob~J. Hughey, Lisa Bastarache, Joshua~C. Denny, and
  Qingxia Chen.
\newblock {Cox regression is robust to inaccurate EHR-extracted event time: an
  application to EHR-based GWAS}.
\newblock {\em Bioinformatics}, 38:2297--2306, March 2022.

\bibitem{wang2016functional}
Jane-Ling Wang, Jeng-Min Chiou, and Hans-Georg M{\"u}ller.
\newblock Functional data analysis.
\newblock {\em Annual Review of Statistics and its application}, 3:257--295,
  2016.

\bibitem{marron2015functional}
James~Stephen Marron, James~O Ramsay, Laura~M Sangalli, and Anuj Srivastava.
\newblock Functional data analysis of amplitude and phase variation.
\newblock {\em Statistical Science}, pages 468--484, 2015.

\bibitem{Ramsay2002CurveR}
J.~O. Ramsay and Xiaochun Li.
\newblock {Curve Registration}.
\newblock {\em Journal of the Royal Statistical Society Series B: Statistical
  Methodology}, 60(2):351--363, 2002.

\bibitem{sakoe1978dynamic}
Hiroaki Sakoe and Seibi Chiba.
\newblock Dynamic programming algorithm optimization for spoken word
  recognition.
\newblock {\em IEEE transactions on acoustics, speech, and signal processing},
  26(1):43--49, 1978.

\bibitem{wang1999synchronizing}
Kongming Wang and Theo Gasser.
\newblock Synchronizing sample curves nonparametrically.
\newblock {\em Annals of Statistics}, pages 439--460, 1999.

\bibitem{lawton1971self}
William~H Lawton and Edward~A Sylvestre.
\newblock Self modeling curve resolution.
\newblock {\em Technometrics}, 13(3):617--633, 1971.

\bibitem{kneip1988convergence}
Alois Kneip and Theo Gasser.
\newblock Convergence and consistency results for self-modeling nonlinear
  regression.
\newblock {\em The Annals of Statistics}, pages 82--112, 1988.

\bibitem{yao2005functional}
Fang Yao, Hans-Georg M{\"u}ller, and Jane-Ling Wang.
\newblock Functional data analysis for sparse longitudinal data.
\newblock {\em Journal of the American statistical association},
  100(470):577--590, 2005.

\bibitem{Kneip2008CombiningRA}
Alois Kneip and James~O. Ramsay.
\newblock {Combining Registration and Fitting for Functional Models}.
\newblock {\em Journal of the American Statistical Association},
  103:1155--1165, 2008.

\bibitem{james2007curve}
Gareth~M James.
\newblock Curve alignment by moments.
\newblock {\em The Annals of Applied Statistics}, 1(2):480, 2007.

\bibitem{sangalli2009case}
Laura~M Sangalli, Piercesare Secchi, Simone Vantini, and Alessandro Veneziani.
\newblock A case study in exploratory functional data analysis: geometrical
  features of the internal carotid artery.
\newblock {\em Journal of the American Statistical Association},
  104(485):37--48, 2009.

\bibitem{srivastava2011registration}
Anuj Srivastava, Wei Wu, Sebastian Kurtek, Eric Klassen, and J.~S. Marron.
\newblock {Registration of Functional Data Using Fisher-Rao Metric}.
\newblock {\em arXiv}, 1103.3817, 2011.

\bibitem{tucker2013generative}
J~Derek Tucker, Wei Wu, and Anuj Srivastava.
\newblock Generative models for functional data using phase and amplitude
  separation.
\newblock {\em Computational Statistics \& Data Analysis}, 61:50--66, 2013.

\bibitem{wu2014analysis}
Wei Wu and Anuj Srivastava.
\newblock {Analysis of spike train data: Alignment and comparisons using the
  extended Fisher-Rao metric}.
\newblock {\em Electronic Journal of Statistics}, 8:1776--1785, 2014.

\bibitem{wrobel2019registration}
Julia Wrobel, Vadim Zipunnikov, Jennifer Schrack, and Jeff Goldsmith.
\newblock Registration for exponential family functional data.
\newblock {\em Biometrics}, 75(1):48--57, 2019.

\bibitem{wrobel2018register}
Julia Wrobel.
\newblock {Register: Registration for exponential family functional data}.
\newblock {\em Journal of Open Source Software}, 3(22):557, 2018.

\bibitem{wrobel2021registr}
Julia Wrobel and Alexander Bauer.
\newblock {registr 2.0: Incomplete curve registration for exponential family
  functional data}.
\newblock {\em Journal of Open Source Software}, 6(61):2964, 2021.

\bibitem{tang2008pairwise}
Rong Tang and Hans-Georg M{\"u}ller.
\newblock Pairwise curve synchronization for functional data.
\newblock {\em Biometrika}, 95(4):875--889, 2008.

\bibitem{peng2014time}
Jie Peng, Debashis Paul, and Hans-Georg M{\"u}ller.
\newblock Time-warped growth processes, with applications to the modeling of
  boom--bust cycles in house prices.
\newblock {\em The Annals of Applied Statistics}, 8:1561--1582, 2014.

\bibitem{mcdonnell2022registration}
Erin~I McDonnell, Vadim Zipunnikov, Jennifer~A Schrack, Jeff Goldsmith, and
  Julia Wrobel.
\newblock Registration of 24-hour accelerometric rest-activity profiles and its
  application to human chronotypes.
\newblock {\em Biological rhythm research}, 53(8):1299--1319, 2022.

\bibitem{hadjipantelis2015unifying}
Pantelis~Z Hadjipantelis, John~AD Aston, Hans-Georg M{\"u}ller, and Jonathan~P
  Evans.
\newblock {Unifying amplitude and phase analysis: A compositional data approach
  to functional multivariate mixed-effects modeling of Mandarin Chinese}.
\newblock {\em Journal of the American Statistical Association},
  110(510):545--559, 2015.

\bibitem{liu2009simultaneous}
Xueli Liu and Mei-Cheng~K. Yang.
\newblock Simultaneous curve registration and clustering for functional data.
\newblock {\em Computational Statistics \& Data Analysis}, 53(4):1361--1376,
  2009.

\bibitem{wu2016bayesian}
Zizhen Wu and David~B. Hitchcock.
\newblock {A Bayesian method for simultaneous registration and clustering of
  functional observations}.
\newblock {\em Computational Statistics \& Data Analysis}, 101:121--136, 2016.

\bibitem{jiang2023timeline}
Shiyi Jiang, Rungang Han, Krishnendu Chakrabarty, David Page, William~W Stead,
  and Anru~R Zhang.
\newblock {Timeline Registration for Electronic Health Records}.
\newblock {\em AMIA Jt Summits Transl Sci Proc}, 2023:291--299, 2023.
\newblock PMID: 37350882; PMCID: PMC10283114.

\bibitem{Goldberger2000PhysionetCO}
Ary~L. Goldberger, Luis A.~Nunes Amaral, L~Glass, Jeffrey~M. Hausdorff,
  Plamen~Ch. Ivanov, Roger~G. Mark, Joseph~E. Mietus, George~B. Moody,
  Chung-Kang Peng, and Harry~Eugene Stanley.
\newblock {PhysioBank, PhysioToolkit, and PhysioNet: components of a new
  research resource for complex physiologic signals}.
\newblock {\em Circulation}, 101(23):E215--20, 2000.

\bibitem{Johnson2023MIMICIVAF}
Alistair E.~W. Johnson, Lucas Bulgarelli, Lu~Shen, Alvin Gayles, Ayad Shammout,
  Steven Horng, Tom~J. Pollard, Benjamin Moody, Brian Gow, Li~wei H.~Lehman,
  Leo~Anthony Celi, and Roger~G. Mark.
\newblock {MIMIC-IV, a freely accessible electronic health record dataset}.
\newblock {\em Scientific Data}, 10:1, 2023.

\bibitem{johnson2023mimicivdemo}
Alistair Johnson, Luca Bulgarelli, Tom Pollard, Steven Horng, Leo~Anthony Celi,
  and Roger Mark.
\newblock {MIMIC-IV Clinical Database Demo (version 2.2)}.
\newblock PhysioNet, 2023.
\newblock RRID:SCR\_007345.

\bibitem{Makris2016AcuteKI}
Konstantinos Makris and Loukia Spanou.
\newblock {Acute Kidney Injury: Definition, Pathophysiology and Clinical
  Phenotypes.}
\newblock {\em The Clinical biochemist. Reviews}, 37(2):85--98, 2016.

\bibitem{Rewa2014AcuteKI}
Oleksa~G. Rewa and Sean~M. Bagshaw.
\newblock Acute kidney injury—epidemiology, outcomes and economics.
\newblock {\em Nature Reviews Nephrology}, 10:193--207, 2014.

\bibitem{song2020cross}
Xing Song, Alan~SL Yu, John~A Kellum, Lemuel~R Waitman, Michael~E Matheny,
  Steven~Q Simpson, Yong Hu, and Mei Liu.
\newblock Cross-site transportability of an explainable artificial intelligence
  model for acute kidney injury prediction.
\newblock {\em Nature Communications}, 11:5668, 2020.

\bibitem{Khwaja2012KDIGOCP}
Arif Khwaja.
\newblock {KDIGO Clinical Practice Guidelines for Acute Kidney Injury}.
\newblock {\em Nephron Clinical Practice}, 120:c179--c184, 2012.

\bibitem{hauskrecht2013outlier}
Milos Hauskrecht, Iyad Batal, Michal Valko, Shyam Visweswaran, Gregory~F
  Cooper, and Gilles Clermont.
\newblock Outlier detection for patient monitoring and alerting.
\newblock {\em Journal of biomedical informatics}, 46(1):47--55, 2013.

\bibitem{wahba1990spline}
Grace Wahba.
\newblock {\em {Spline Models for Observational Data}}.
\newblock CBMS-NSF Regional Conference Series in Applied Mathematics. Society
  for Industrial and Applied Mathematics, Philadelphia, PA, 1990.

\bibitem{hastie2009elements}
Trevor Hastie, Robert Tibshirani, Jerome~H Friedman, and Jerome~H Friedman.
\newblock {\em The elements of statistical learning: data mining, inference,
  and prediction}, volume~2.
\newblock Springer, 2009.

\bibitem{wahba2014spline}
Grace Wahba and Yuedong Wang.
\newblock {Spline Functions: Overview}.
\newblock {\em Wiley StatsRef: Statistics Reference Online}, pages 1--19, 2014.

\bibitem{rousseeuw1987silhouettes}
Peter~J Rousseeuw.
\newblock Silhouettes: a graphical aid to the interpretation and validation of
  cluster analysis.
\newblock {\em Journal of computational and applied mathematics}, 20:53--65,
  1987.

\bibitem{macqueen1967some}
James MacQueen.
\newblock Some methods for classification and analysis of multivariate
  observations.
\newblock In {\em Proceedings of the Fifth Berkeley Symposium on Mathematical
  Statistics and Probability}, volume~1, pages 281--297. Oakland, CA, USA,
  1967.

\bibitem{rdusseeun1987clustering}
LKPJ Rdusseeun and P~Kaufman.
\newblock Clustering by means of medoids.
\newblock In {\em Proceedings of the statistical data analysis based on the L1
  norm conference, neuchatel, switzerland}, volume~31, 1987.

\bibitem{Zhu2024FunctionalPS}
Zihan Zhu, Xin Gai, and Anru~R. Zhang.
\newblock {Functional Post-Clustering Selective Inference with Applications to
  EHR Data Analysis}.
\newblock {\em arXiv}, 2405.03042, 2024.

\bibitem{Levey1999AMA}
Andrew~S. Levey, Juan~Ponsa Bosch, Julia~Breyer Lewis, Tom Greene, Nancy~L.
  Rogers, and David Roth.
\newblock {A More Accurate Method To Estimate Glomerular Filtration Rate from
  Serum Creatinine: A New Prediction Equation}.
\newblock {\em Annals of Internal Medicine}, 130:461--470, 1999.

\bibitem{Jiang2023SoftPF}
Shiyi Jiang, Xin Gai, Miriam~M. Treggiari, William~W. Stead, Yuankang Zhao,
  David Page, and Anru~R. Zhang.
\newblock {Soft phenotyping for sepsis via EHR time-aware soft clustering}.
\newblock {\em Journal of Biomedical Informatics}, 152:104615, 2023.

\bibitem{Lipton2016LearningTD}
Zachary~Chase Lipton, David~C. Kale, Charles~Peter Elkan, and Randall~C.
  Wetzel.
\newblock {Learning to Diagnose with LSTM Recurrent Neural Networks}.
\newblock In {\em ICLR}, 2016.

\bibitem{matthay2020phenotypes}
Michael~A Matthay, Yaseen~M Arabi, Emily~R Siegel, Lorraine~B Ware, Lieuwe D~J
  Bos, Pratik Sinha, Jeremy~R Beitler, Katherine~D Wick, Martha A~Q Curley,
  Jean-Michel Constantin, Joseph~E Levitt, and Carolyn~S Calfee.
\newblock Phenotypes and personalized medicine in the acute respiratory
  distress syndrome.
\newblock {\em Intensive Care Medicine}, 46(12):2136--2152, 2020.

\bibitem{wick2021promises}
Katherine~D Wick, Daniel~F McAuley, Joseph~E Levitt, Jeremy~R Beitler, Djillali
  Annane, Elisabeth~D Riviello, Carolyn~S Calfee, and Michael~A Matthay.
\newblock {Promises and challenges of personalized medicine to guide ARDS
  therapy}.
\newblock {\em Critical Care}, 25(1):404, 2021.

\bibitem{mcdonald2019treatable}
Vanessa~M McDonald, James Fingleton, Alvar Agusti, Sarah~A Hiles, Vanessa~L
  Clark, Anne~E Holland, Guy~B Marks, Philip~P Bardin, Richard Beasley, Ian~D
  Pavord, Peter A~B Wark, and Peter~G Gibson.
\newblock {Treatable traits: a new paradigm for 21st century management of
  chronic airway diseases: Treatable Traits Down Under International Workshop
  report}.
\newblock {\em European Respiratory Journal}, 53(5):1802058, 2019.

\bibitem{zarbock2023sepsis}
Alexander Zarbock, Mitra~K. Nadim, Peter Pickkers, Hernando Gomez, Samira Bell,
  Michael Joannidis, Kianoush Kashani, Jay~L. Koyner, Neesh Pannu, Melanie
  Meersch, Thiago Reis, Thomas Rimmelé, Sean~M. Bagshaw, Rinaldo Bellomo,
  Vicenzo Cantaluppi, Akash Deep, Silvia De~Rosa, Xose Perez-Fernandez, Faeq
  Husain-Syed, Sandra~L. Kane-Gill, Yvelynne Kelly, Ravindra~L. Mehta,
  Patrick~T. Murray, Marlies Ostermann, John Prowle, Zaccaria Ricci, Emily~J.
  See, Antoine Schneider, Danielle~E. Soranno, Ashita Tolwani, Gianluca Villa,
  Claudio Ronco, and Lui~G. Forni.
\newblock {Sepsis-associated acute kidney injury: consensus report of the 28th
  Acute Disease Quality Initiative workgroup}.
\newblock {\em Nature Reviews Nephrology}, 19(6):401--417, 2023.

\bibitem{rahman2012acute}
Mahboob Rahman, Fariha Shad, and Michael~C. Smith.
\newblock Acute kidney injury: a guide to diagnosis and management.
\newblock {\em American Family Physician}, 86(7):631--639, 2012.

\bibitem{rangaswami2019cardiorenal}
Janani Rangaswami, Vivek Bhalla, John~E.A. Blair, Tara~I. Chang, Salvatore
  Costa, Krista~L. Lentine, Edgar~V. Lerma, Kenechukwu Mezue, Mark Molitch,
  Wilfried Mullens, Claudio Ronco, W.H.~Wilson Tang, and Peter~A. McCullough.
\newblock {Cardiorenal Syndrome: Classification, Pathophysiology, Diagnosis,
  and Treatment Strategies: A Scientific Statement From the American Heart
  Association}.
\newblock {\em Circulation}, 139(16):e840--e878, 2019.

\bibitem{HusainSyed2016LungKidney}
Faeq Husain-Syed, Arthur~S. Slutsky, and Claudio Ronco.
\newblock {Lung--Kidney Cross-Talk in the Critically Ill Patient}.
\newblock {\em American Journal of Respiratory and Critical Care Medicine},
  194(4):402--414, 2016.

\bibitem{Xu2021SystematicRA}
Zhenjian Xu, Ying Tang, Qiuyan Huang, Sha Fu, Xiaomei Li, Baojuan Lin, Anping
  Xu, and Junzhe Chen.
\newblock {Systematic review and subgroup analysis of the incidence of acute
  kidney injury (AKI) in patients with COVID-19}.
\newblock {\em BMC Nephrology}, 22(1):52, 2021.

\bibitem{zhang2025development}
Yuhui Zhang, Damin Xu, Jianwei Gao, Ruiguo Wang, Kun Yan, Hong Liang, Juan Xu,
  Youlu Zhao, Xizi Zheng, Lingyi Xu, Jinwei Wang, Fude Zhou, Guopeng Zhou,
  Qingqing Zhou, Zhao Yang, Xiaoli Chen, Yulan Shen, Tianrong Ji, Yunlin Feng,
  Ping Wang, Jundong Jiao, Li~Wang, Jicheng Lv, and Li~Yang.
\newblock Development and validation of a real-time prediction model for acute
  kidney injury in hospitalized patients.
\newblock {\em Nature Communications}, 16(1):68, 2025.

\bibitem{Muller2007}
Meinard M{\"u}ller.
\newblock {Dynamic Time Warping}.
\newblock In {\em Information Retrieval for Music and Motion}, pages 69--84.
  Springer Berlin Heidelberg, Berlin, Heidelberg, 2007.

\bibitem{luo2024adaptive}
Yuetian Luo and Chao Gao.
\newblock {Adaptive Robust Confidence Intervals}.
\newblock {\em arXiv preprint arXiv:2410.22647}, 2024.

\bibitem{dalmaijer2022statistical}
Edwin~S Dalmaijer, Camilla~L Nord, and Duncan~E Astle.
\newblock Statistical power for cluster analysis.
\newblock {\em BMC Bioinformatics}, 23(1):205, 2022.

\bibitem{banerjee2023identifying}
Amitava Banerjee, Arezoo Dashtban, Sheng Chen, Rosita Zakeri, Priyanka Gupta,
  Laura Pasea, Maria Thomas, Andrew Wood, Nadia Raihan, Ali Mulla, Nyein Aung,
  Spiros Denaxas, AD~Hingorani, and Harry Hemingway.
\newblock Identifying subtypes of heart failure from three electronic health
  record sources with machine learning: an external, prognostic, and genetic
  validation study.
\newblock {\em The Lancet Digital Health}, 5(6):e370--e379, 2023.

\bibitem{manzini2022longitudinal}
Elisa Manzini, Blerina Vlacho, Josep Franch-Nadal, Mercè Fernández, Manel
  Mata-Cases, Dídac Mauricio, Núria Alcubierre, Jordi Real, María~Eugenia
  Mauricio, and Iñaki Cano.
\newblock {Longitudinal deep learning clustering of Type 2 Diabetes Mellitus
  trajectories using routinely collected health records}.
\newblock {\em Journal of Biomedical Informatics}, 135:104218, 2022.

\end{thebibliography}

\end{sloppypar}

\newpage
\appendix

\setcounter{page}{1}

\begin{sloppypar}
	
	\begin{center}
		\Large{Supplementary Materials for ``Subtype-Aware Registration of Longitudinal Electronic Health Records"}
	\end{center}
	
	\begin{abstract}
		We provide simulation studies in these supplementary materials.
	\end{abstract}
	
	\section{Simulations}\label{sec:simulation}
	
	In this section, we conduct simulation studies across eight scenarios to evaluate the ability of the proposed algorithm to recover the true temporal positions of longitudinal trajectories under diverse shapes, mean structures, and progression speeds. We first describe the common setup, then the specific designs, and finally the results.
	\subsection{Common Simulation Setup} \label{sec:simusetup}
	
	\textit{Data Generation:} For each subject \( i \) (where \( i = 1, 2, \ldots, 1000 \)), the observation time points \( t_{ij} \) (where \( j = 1, 2, \ldots, 28 \)) were randomly drawn from a continuous uniform distribution over the interval $[1, 21]$ (\( t_{ij} \sim \text{Uniform}(1, 21) \)). These time points were then sorted in ascending order for each subject. In all scenarios, the noise terms \( \epsilon_{ij} \) were randomly drawn from a normal distribution \( \epsilon_{ij} \sim N(0, 0.8^2) \).
	
	\textit{Temporal Misalignment Simulation:} In all scenarios, each subject was assigned a time shift to simulate the delay between the onset of illness and hospital admission. Specifically, in each scenario, 60\% of the subjects in a group had a time shift of 0, representing patients who were admitted to the hospital immediately upon illness onset or were promptly transferred from general wards to the ICU after the onset of severe symptoms. The remaining 40\% of subjects were assigned a random time shift ranging from 1 to 4 days (integers), simulating a delay of 1 to 4 days between illness onset and hospital admission. After applying the time shifts, any observations with time points less than 1 were removed to simulate the missing data during the ``onset-to-admission" period when patients experience a delay in hospital admission. Additionally, observations with time points greater than 17 were removed to ensure a consistent study time window across all individuals.
	
	\textit{Algorithm Parameter Settings:} We fix 2 internal knots at 8 and 13. The potential shift positions \(\Lambda\) were set to \(\{0, 1, 2, 3, 4\}\). The range of cluster numbers was set to \(\{2, 3, \ldots, 8\}\), i.e., \( M = 8 \). The selection proportion \(\alpha\) was set to 0.95, and the threshold \(\tau\) was set to 0.45. 
	We employ $k$-medoids as the clustering method employed given its robustness to outliers \cite{rdusseeun1987clustering}.

	\subsection{Scenario Design}
	Eight scenarios are constructed to represent different combinations of trajectory characteristics:
	
	\begin{itemize}
		\item \textbf{Scenarios 1--2:} Distinguish trajectories based on either different means or different shapes.
		\item \textbf{Scenarios 3--4:} Incorporate both shape and mean differences with overlapping patterns and increased cluster complexity.
		\item \textbf{Scenarios 5--6:} Add intra-shape variation in mean levels and increased inter-group similarity.
		\item \textbf{Scenarios 7--8:} Introduce within-group progression speed heterogeneity via compressed/dilated time scales.
	\end{itemize}
	
	For completeness, we provide the trajectory-generating equations used in these eight scenarios. Each trajectory includes an additive Gaussian noise term \(\epsilon_{ij} \sim N(0, 0.8^2)\). Heterogeneity in progression speed is encoded via a subject-specific factor \(s_i^{(\cdot)}\) where applicable.
	
	\paragraph{Scenario 1:} Two groups with similar shapes but different means:
	\begin{align*}
		\text{Group 1 (500 subjects):} \quad Y_{i}^{(1)}(t_{ij}) &= 20 + 3 \sin(0.6(t_{ij} + 4)) + \epsilon_{ij}, \\
		\text{Group 2 (500 subjects):} \quad Y_{i}^{(2)}(t_{ij}) &= 17 + 3 \sin(0.6(t_{ij} + 4)) + \epsilon_{ij}.
	\end{align*}
	
	\paragraph{Scenario 2:} Two groups with similar means but different shapes:
	\begin{align*}
		\text{Group 1 (500 subjects):} \quad Y_{i}^{(1)}(t_{ij}) &= 17 + 2 \sin(0.6(t_{ij} + 4)) + \epsilon_{ij}, \\
		\text{Group 2 (500 subjects):} \quad Y_{i}^{(2)}(t_{ij}) &= 2 t_{ij} + 3 \sin(0.6(t_{ij} + 4)) + \epsilon_{ij}.
	\end{align*}
	
	\paragraph{Scenario 3:} Two groups with shared shapes and different means; a third group with a different shape:
	\begin{align*}
		\text{Group 1 (300 subjects):} \quad Y_{i}^{(1)}(t_{ij}) &= 20 + 3 \sin(0.6(t_{ij} + 4)) + \epsilon_{ij}, \\
		\text{Group 2 (400 subjects):} \quad Y_{i}^{(2)}(t_{ij}) &= 17 + 3 \sin(0.6(t_{ij} + 4)) + \epsilon_{ij}, \\
		\text{Group 3 (300 subjects):} \quad Y_{i}^{(3)}(t_{ij}) &= 16 + 0.5 t_{ij} + 3 \sin(0.9(t_{ij} + 4)) + \epsilon_{ij}.
	\end{align*}
	
	\paragraph{Scenario 4:} Four groups with overlapping shapes and means:
	\begin{align*}
		\text{Group 1 (250 subjects):} \quad Y_{i}^{(1)}(t_{ij}) &= 24 + 3 \sin(0.6(t_{ij} + 4)) + \epsilon_{ij}, \\
		\text{Group 2 (250 subjects):} \quad Y_{i}^{(2)}(t_{ij}) &= 17 + 3 \sin(0.6(t_{ij} + 4)) + \epsilon_{ij}, \\
		\text{Group 3 (250 subjects):} \quad Y_{i}^{(3)}(t_{ij}) &= 38 - 0.5 t_{ij} + 4 \sin(0.9(t_{ij} + 3)) + \epsilon_{ij}, \\
		\text{Group 4 (250 subjects):} \quad Y_{i}^{(4)}(t_{ij}) &= 21 + 0.9 t_{ij} + 3 \sin(0.6(t_{ij} + 4)) + \epsilon_{ij}.
	\end{align*}
	
	\paragraph{Scenario 5:} Four groups with two shared shapes and different means:
	\begin{align*}
		\text{Group 1 (250 subjects):} \quad Y_{i}^{(1)}(t_{ij}) &= 24 + 3 \sin(0.6(t_{ij} + 4)) + \epsilon_{ij}, \\
		\text{Group 2 (250 subjects):} \quad Y_{i}^{(2)}(t_{ij}) &= 17 + 3 \sin(0.6(t_{ij} + 4)) + \epsilon_{ij}, \\
		\text{Group 3 (250 subjects):} \quad Y_{i}^{(3)}(t_{ij}) &= 28 - 0.5 t_{ij} + 4 \sin(0.9(t_{ij} + 3)) + \epsilon_{ij}, \\
		\text{Group 4 (250 subjects):} \quad Y_{i}^{(4)}(t_{ij}) &= 21 - 0.5 t_{ij} + 4 \sin(0.9(t_{ij} + 3)) + \epsilon_{ij}.
	\end{align*}
	
	\paragraph{Scenario 6:} Three groups with a shared shape and one group with a distinct shape:
	\begin{align*}
		\text{Group 1 (250 subjects):} \quad Y_{i}^{(1)}(t_{ij}) &= 23 + 3 \sin(0.6(t_{ij} + 4)) + \epsilon_{ij}, \\
		\text{Group 2 (250 subjects):} \quad Y_{i}^{(2)}(t_{ij}) &= 17 + 3 \sin(0.6(t_{ij} + 4)) + \epsilon_{ij}, \\
		\text{Group 3 (250 subjects):} \quad Y_{i}^{(3)}(t_{ij}) &= 20 + 3 \sin(0.6(t_{ij} + 4)) + \epsilon_{ij}, \\
		\text{Group 4 (250 subjects):} \quad Y_{i}^{(4)}(t_{ij}) &= 25 - 0.5 t_{ij} + 4 \sin(0.9(t_{ij} + 3)) + \epsilon_{ij}.
	\end{align*}
	
	\paragraph{Scenario 7:} Heterogeneous progression speeds within Groups~2 and 3, with $s_i^{(2a)}=1$ and $s_i^{(2b)}=0.7$:
	\begin{align*}
		\text{Group 1 (500 subjects):}\quad Y_{i}^{(1)}(t_{ij}) &= 17 + 2 \sin\!\big(0.6(t_{ij} + 4)\big) + \epsilon_{ij}, \\
		\text{Group 2 (250 subjects):}\quad 
		Y_{i}^{(2a)}\!\big(s_i^{(2a)} t_{ij}\big) &= 2\,t_{ij} + 3 \sin\!\big(0.6(t_{ij} + 4)\big) + \epsilon_{ij}, \\
		\text{Group 3 (250 subjects):}\quad 
		Y_{i}^{(2b)}\!\big(s_i^{(2b)} t_{ij}\big) &= 2\,t_{ij} + 3 \sin\!\big(0.6(t_{ij} + 4)\big) + \epsilon_{ij}.
	\end{align*}
	
	\paragraph{Scenario 8:} Heterogeneous progression speeds across Groups~3 and 4, with $s_i^{(3a)}=1$ and $s_i^{(3b)}=1.3$:
	\begin{align*}
		\text{Group 1 (300 subjects):}\quad Y_{i}^{(1)}(t_{ij}) &= 20 + 3 \sin\!\big(0.6(t_{ij} + 4)\big) + \epsilon_{ij}, \\
		\text{Group 2 (300 subjects):}\quad 
		Y_{i}^{(2)}(t_{ij}) &= 17 + 3 \sin\!\big(0.6(t_{ij} + 4)\big) + \epsilon_{ij}, \\
		\text{Group 3 (200 subjects):}\quad 
		Y_{i}^{(3a)}\!\big(s_i^{(3a)} t_{ij}\big) &= 16 + 0.5\,t_{ij} + 3 \sin\!\big(0.9(t_{ij} + 4)\big) + \epsilon_{ij}, \\
		\text{Group 4 (200 subjects):}\quad 
		Y_{i}^{(3b)}\!\big(s_i^{(3b)} t_{ij}\big) &= 16 + 0.5\,t_{ij} + 3 \sin\!\big(0.9(t_{ij} + 4)\big) + \epsilon_{ij}.
	\end{align*}
	
\begin{figure}[hbt!]
		\centering
		\includegraphics[width=.90\textwidth]{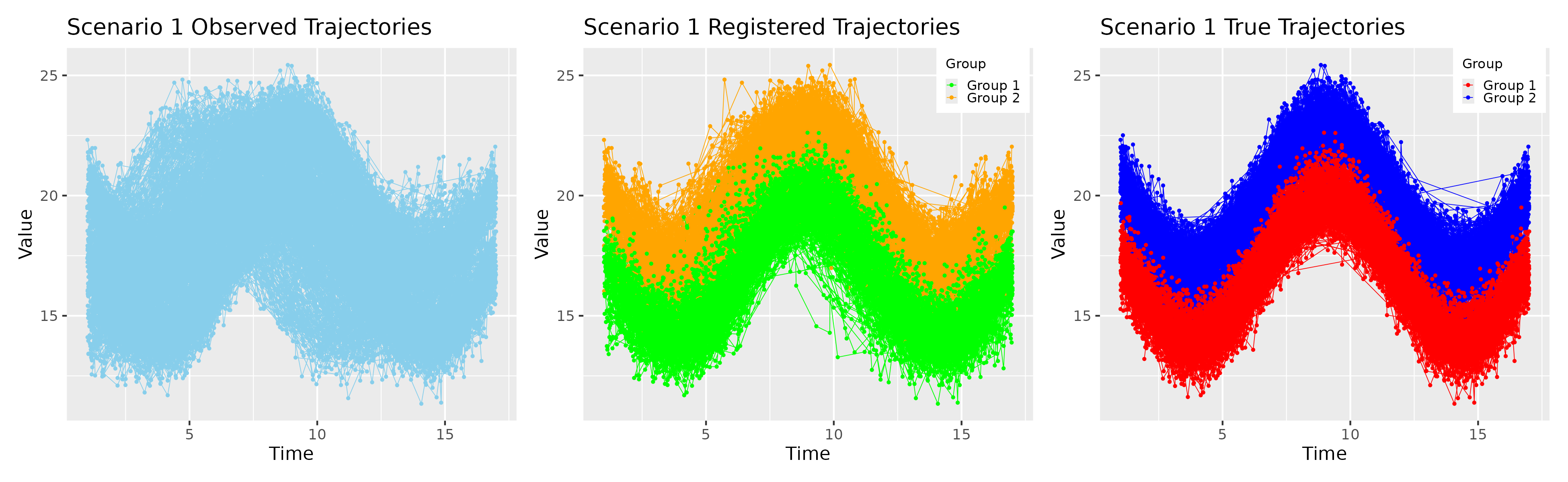}
		\caption[Trajectories of Simu1.]{Simulation 1 Trajectories: Left: Observed trajectories with unregistered timelines and unknown subtypes. Right: True trajectories with subtypes indicated by colors. Middle: Trajectories after applying the proposed timeline registration, with recovered subtypes marked by colors.}
		\label{fig:S1}
	\end{figure}

	An illustrative example for Scenario 1 is shown in Figure~\ref{fig:S1}. Additional visualizations for all scenarios are provided in Figures \ref{fig:S2}–\ref{fig:S8}.

	\subsection{Results}
	We use four metrics to evaluate the performance of the algorithm: Accuracy Rate refers to the proportion of subjects for whom the algorithm successfully recovered the correct temporal positions of their longitudinal data; Probability of Error \(\leq 1\) Day refers to the proportion of subjects with a result error of less than or equal to 1 day; Mean Absolute Error (MAE) represents the average time difference between the recovered temporal positions and the correct temporal positions for all subjects, and the Runtime indicates the time taken for the algorithm to run once on the simulated dataset. The simulation results, which present the average values and 95\% confidence intervals for all four metrics over 500 runs, are shown in Table \ref{tab:Simulation_Results}. The visualizations of the simulation results can be found in Figure \ref{fig:S1} and Figures \ref{fig:S2}, \ref{fig:S3}, \ref{fig:S4}, \ref{fig:S5}, \ref{fig:S6}, \ref{fig:S7}, and \ref{fig:S8} in the supplementary materials.
	
	\begin{table}[!ht]
		\centering
		\resizebox{\textwidth}{!}{%
			\begin{tabular}{c|c|c|c|c}
				\hline
				\textbf{Scenario} & \textbf{Rate of Exact Recovery} & \textbf{Rate of Recovery Error $\leq$ 1} & \textbf{MAE (days)} & \textbf{Runtime (min)} \\ \hline
				1 (2 clusters) & 65.2\% [61.7\%, 74.2\%] & 83.4\% [76.5\%, 97.6\%] & 0.56 [0.29, 0.69] & 0.77 [0.61, 1.11] \\ \hline
				2 (2 clusters) & 75.5\% [71.8\%, 79.1\%] & 93.5\% [89.7\%, 96.4\%] & 0.37 [0.27, 0.47] & 0.72 [0.57, 1.19] \\ \hline
				3 (3 clusters) & 78.1\% [76.3\%, 81.4\%] & 91.2\% [90.2\%, 92.8\%] & 0.36 [0.31, 0.39] & 0.69 [0.61, 0.81] \\ \hline
				4 (4 clusters) & 73.4\% [70.7\%, 82.9\%] & 96.9\% [95.6\%, 98.2\%] & 0.33 [0.25, 0.38] & 0.71 [0.56, 1.19] \\ \hline
				5 (4 clusters) & 70.8\% [60.3\%, 79.4\%] & 90.9\% [87.9\%, 93.5\%] & 0.54 [0.39, 0.68] & 1.13 [0.62, 1.94] \\ \hline
				6 (4 clusters) & 79.7\% [74.7\%, 83.6\%] & 93.9\% [89.8\%, 96.4\%] & 0.37 [0.27, 0.53] & 0.72 [0.57, 1.19] \\ \hline
				7 (3 clusters) & 72.4\% [69.4\%, 76.1\%] & 92.7\% [91.7\%, 94.5\%] & 0.43 [0.37, 0.48] & 0.73 [0.58, 1.21] \\ \hline
				8 (3 clusters) & 71.9\% [60.9\%, 75.9\%] & 86.8\% [79.7\%, 89.3\%] & 0.50 [0.42, 0.71] & 0.74 [0.64, 0.93] \\ \hline
			\end{tabular}%
		}
		\caption{Simulation Results: We present the average value and 95\% confidence interval for the rate of exact recovery, rate of recovery error, mean absolute value, and runtime}
		\label{tab:Simulation_Results}
	\end{table}
	
	\begin{figure}[!hbt]
		\centering
		\includegraphics[width=.99\textwidth]{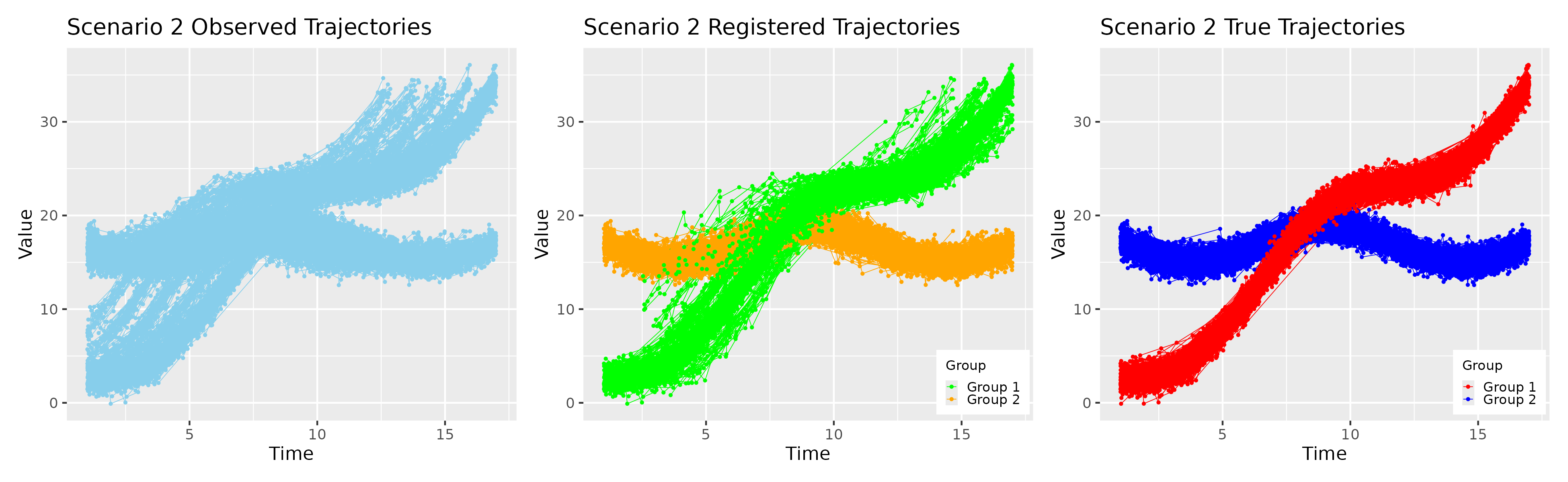}
		\caption[Trajectories of Simu2.]{Simulation 2 trajectories}
		\label{fig:S2}
	\end{figure}
	
	\begin{figure}[!hbt]
		\centering
		\includegraphics[width=.99\textwidth]{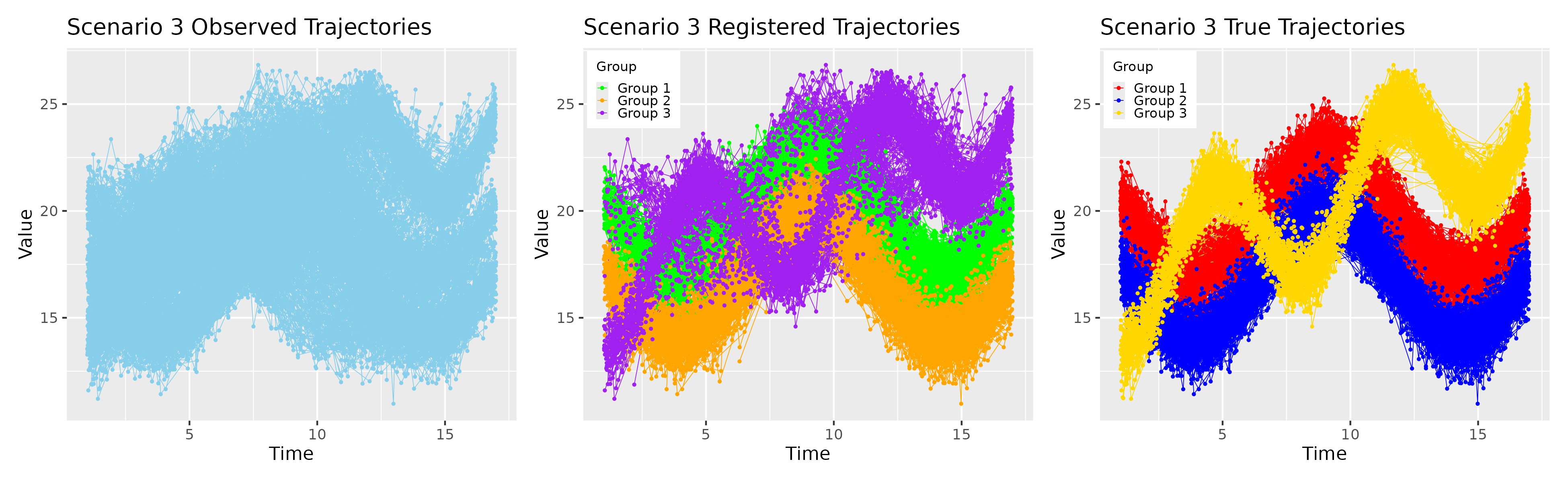}
		\caption[Trajectories of Simu3.]{Simulation 3 trajectories}
		\label{fig:S3}
	\end{figure}
	
	\begin{figure}[!hbt]
		\centering
		\includegraphics[width=.99\textwidth]{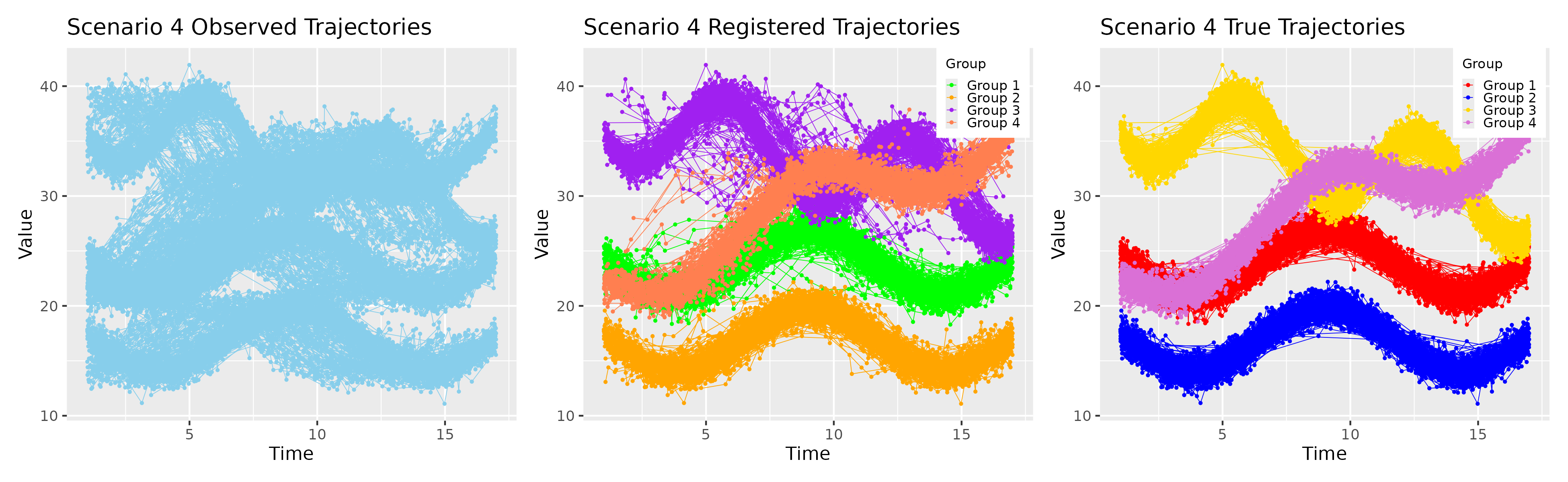}
		\caption[Trajectories of Simu4.]{Simulation 4 trajectories}
		\label{fig:S4}
	\end{figure}
	
	\begin{figure}[!hbt]
		\centering
		\includegraphics[width=.99\textwidth]{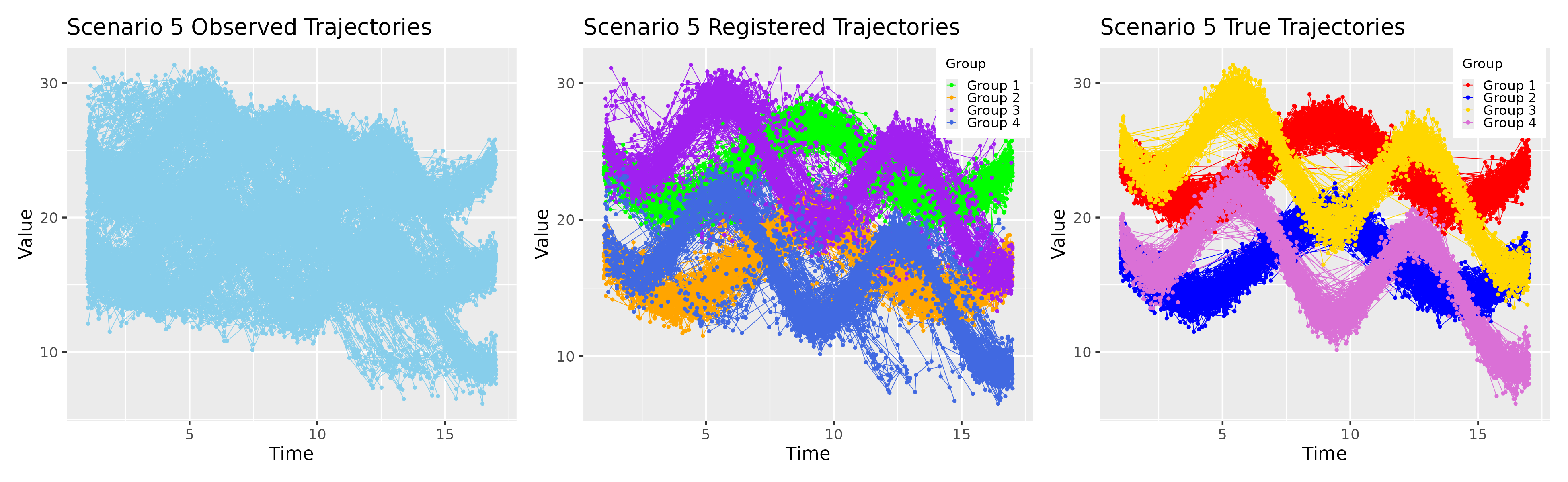}
		\caption[Trajectories of Simu5.]{Simulation 5 trajectories}
		\label{fig:S5}
	\end{figure}
	
	\begin{figure}[!hbt]
		\centering
		\includegraphics[width=.99\textwidth]{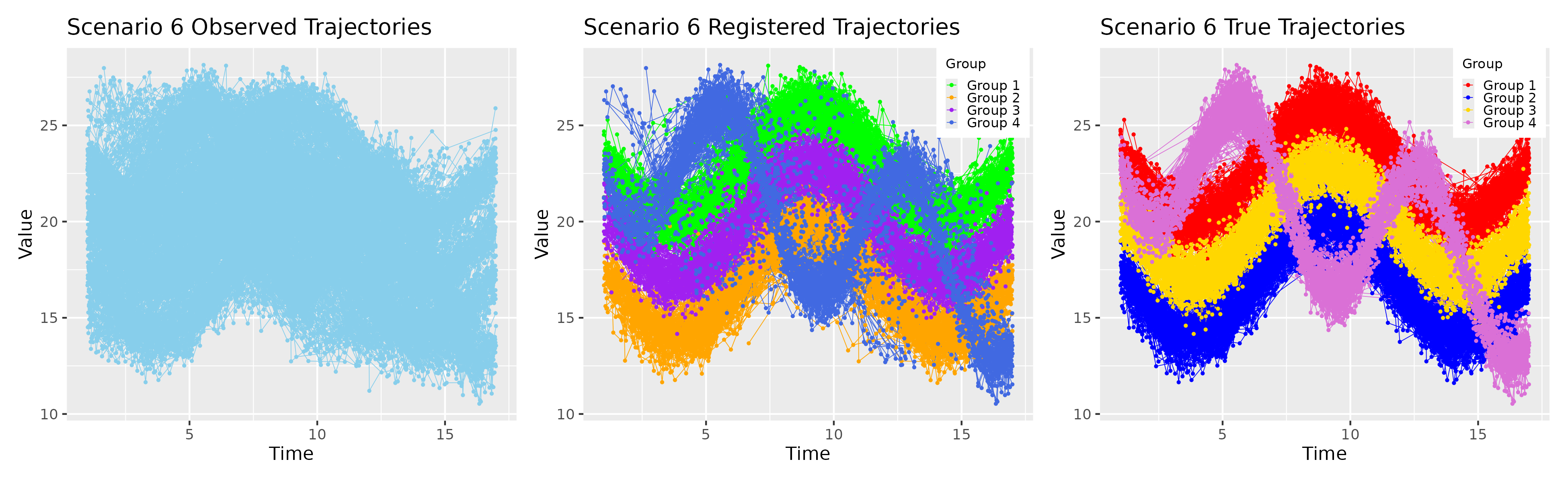}
		\caption[Trajectories of Simu6.]{Simulation 6 trajectories}
		\label{fig:S6}
	\end{figure}
	
	\begin{figure}[!hbt]
		\centering
		\includegraphics[width=.99\textwidth]{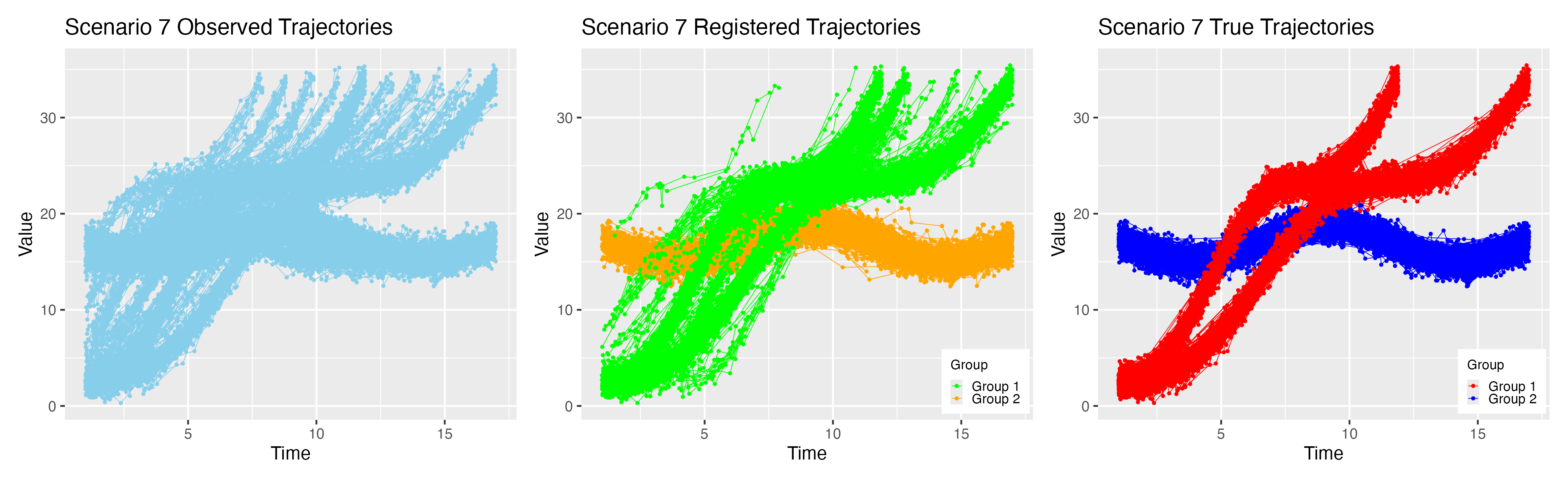}
		\caption[Trajectories of Simu7.]{Simulation 7 trajectories}
		\label{fig:S7}
	\end{figure}
	
	\begin{figure}[!hbt]
		\centering
		\includegraphics[width=.99\textwidth]{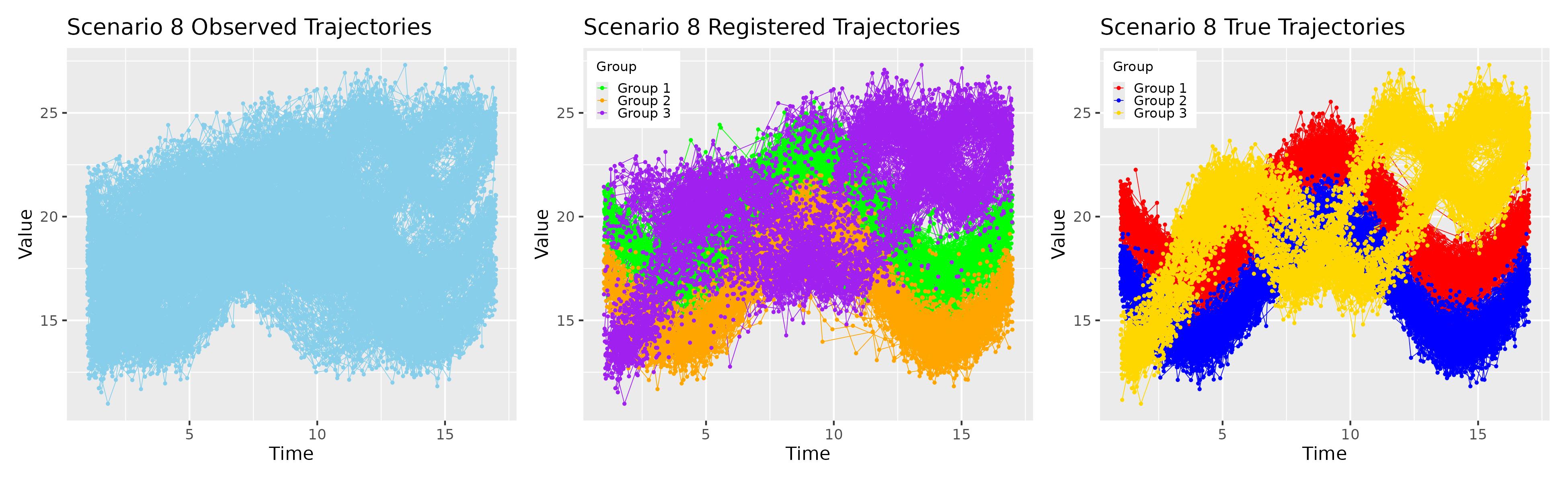}
		\caption[Trajectories of Simu8.]{Simulation 8 trajectories}
		\label{fig:S8}
	\end{figure}

	In summary, our algorithm effectively recovered the temporal positions from the simulated data across all eight scenarios. The evaluation metrics show similar values regardless of the complexity of the scenario, indicating the robustness of our algorithm. The total runtime per execution of the algorithm on the simulated dataset does not exceed 2 minutes, demonstrating its low computational complexity and potential for practical application on large-scale EHR datasets. We further compared our approach with dynamic time warping (DTW) \cite{Muller2007} under heterogeneous progression speeds. The results, presented in Section \ref{appendix:CDTW} of the Supplementary Materials, show the applicability of our method in a broader range of settings.
	
	\subsection{Sensitivity Analyses} \label{sec:SA}
	
\begin{figure}[!hbt]
		\centering
		\includegraphics[width=.9\textwidth]{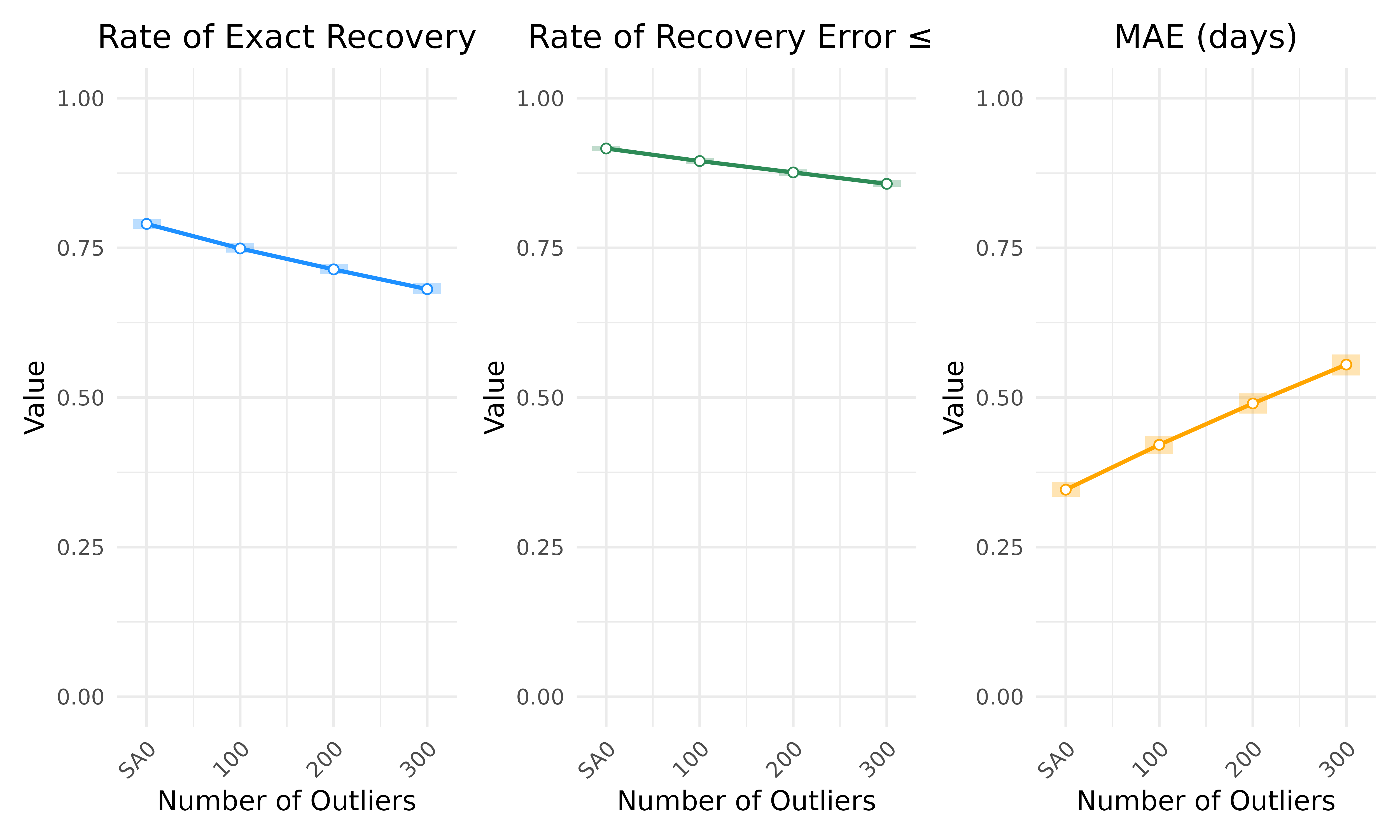}
		\caption[SA 1.]{Validation of robustness to outliers. SA0 represents the original Scenario 3; the other settings were generated from SA0 by randomly selecting 100, 200, and 300 observations, respectively, multiplying them by 2, and thereby introducing progressively higher levels of outlier contamination. In each panel, the shaded band denotes the interquartile range (25th–75th percentiles) of the performance metric, and the central dot marks the median.}
		\label{fig:SA1}
	\end{figure}

\begin{figure}[!hbt]
		\centering
		\includegraphics[width=.9\textwidth]{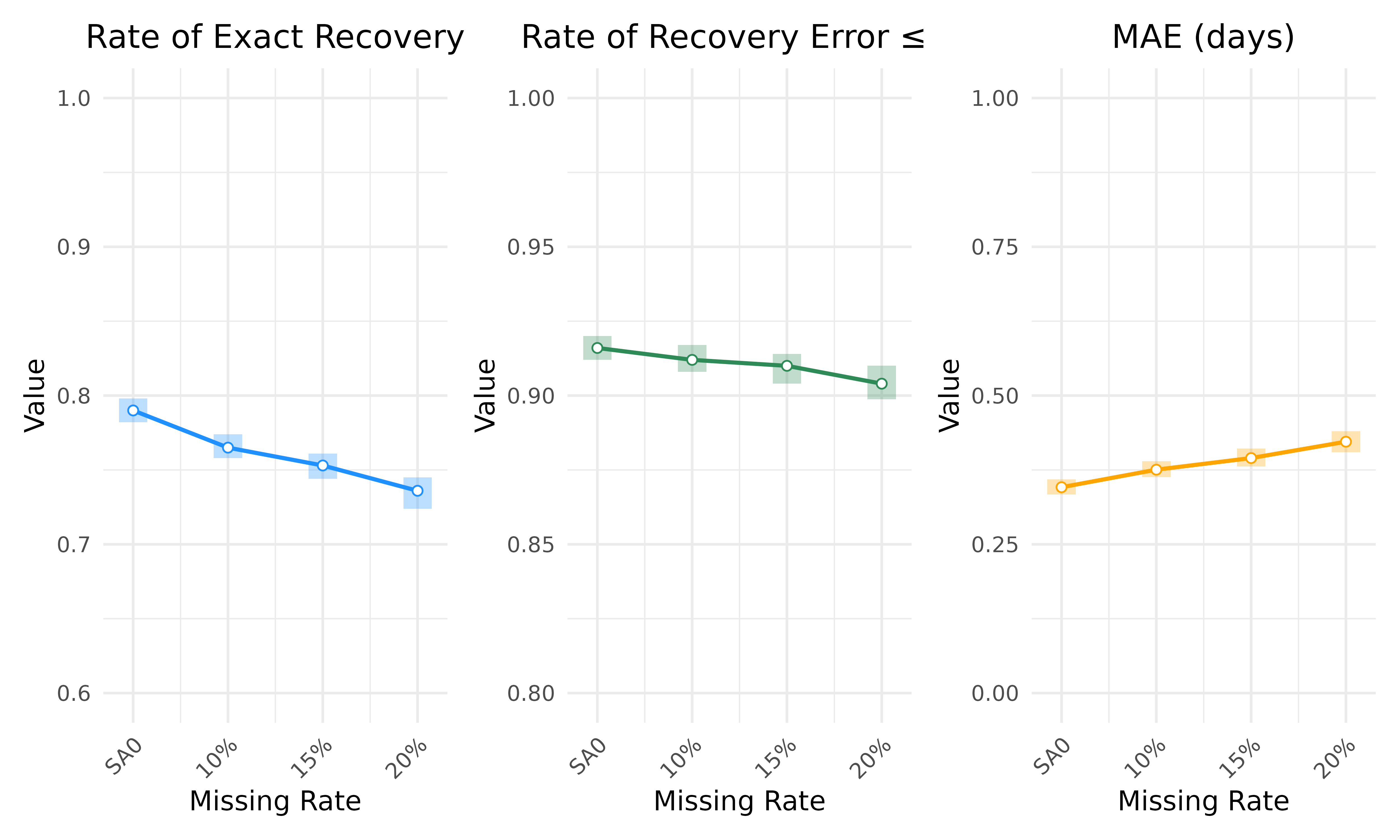}
		\caption[SA 2.]{Validation of robustness to missing data. SA0 denotes the original Scenario 3; the other settings were created from SA0 during data generation by randomly deleting 10\%, 15\%, and 20\% of the generated data points, thus introducing increasing degrees of missingness. In each panel, the shaded band denotes the interquartile range (25th–75th percentiles) of the performance metric, and the central dot marks the median.}
		\label{fig:SA2}
	\end{figure}

\begin{figure}[!hbt]
		\centering
		\includegraphics[width=.9\textwidth]{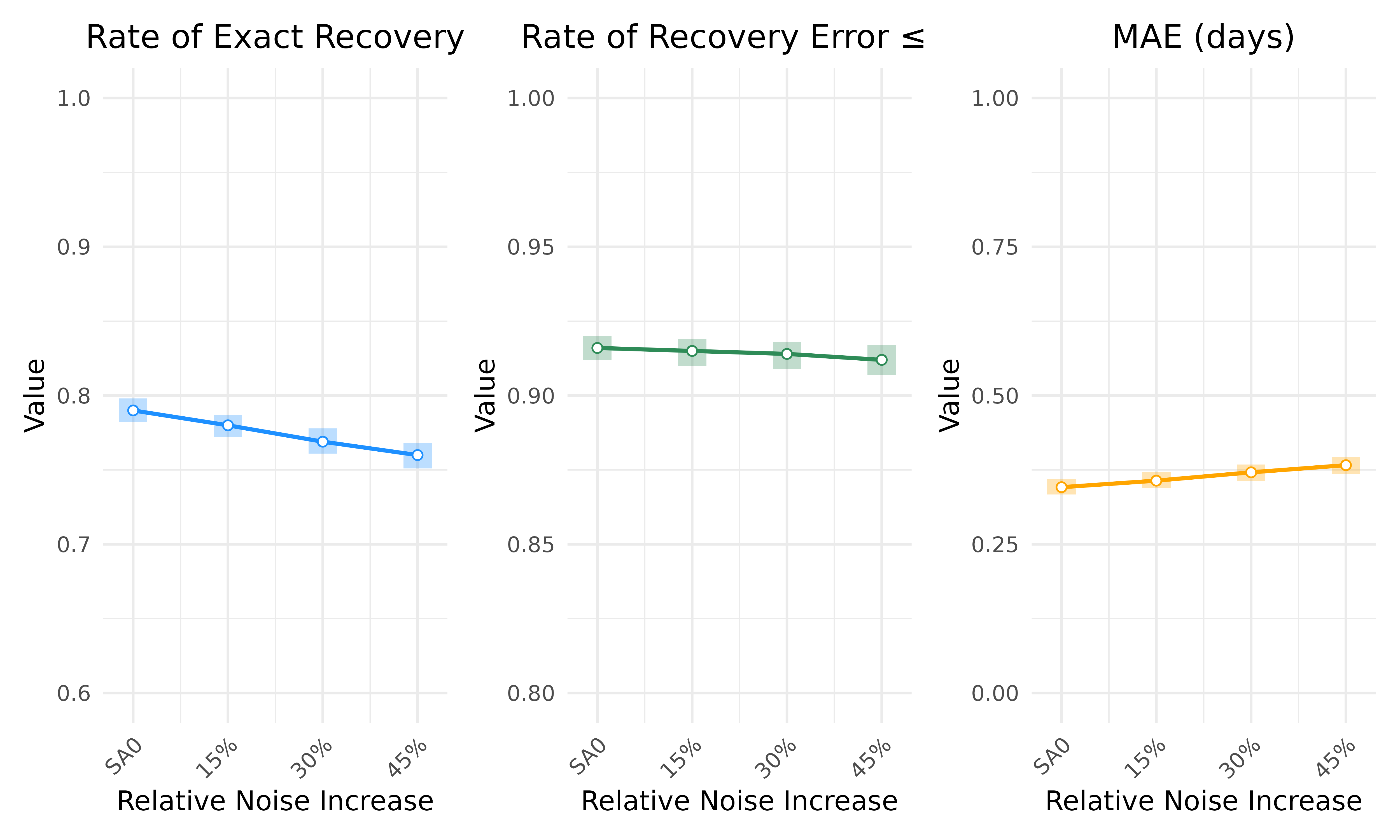}
		\caption[SA 3.]{Validation of robustness to noise. SA0 corresponds to the original Scenario 3; the other settings were generated from SA0 by increasing the variance of the Gaussian error term~$\epsilon_{ij}$ by 15\%, 30\%, and 45\%, respectively, thereby introducing higher noise levels. In each panel, the shaded band denotes the interquartile range (25th–75th percentiles) of the performance metric, and the central dot marks the median.}
		\label{fig:SA3}
	\end{figure}

	Next, we assess the robustness of the proposed algorithm to common challenges in real-world EHR data, including outliers, missingness, and measurement noise. Here, sparsity and irregularity were modeled by randomly sampling observation time points from a uniform distribution over the interval $[1, 21]$, with non-selected times considered missing. Noise was introduced using a random error term, $\epsilon_{ij}$, added to each simulated trajectory. We evaluate: (1) robustness to outliers by modifying the original simulation (Scenario 3, denoted SA0) to randomly select and double the values of 100, 200, and 300 observations, thereby introducing increasing levels of contamination (Figure~\ref{fig:SA1}); (2) robustness to missing data by randomly deleting 10\%, 15\%, and 20\% of the data points at the generation stage (Figure~\ref{fig:SA2}); and (3) robustness to noise by increasing the variance of the additive Gaussian error term $\epsilon_{ij}$ by 15\%, 30\%, and 45\%, respectively (Figure~\ref{fig:SA3}). Across all settings, the algorithm demonstrated stable performance. 
	
	Moreover, we conducted systematic sensitivity analyses to examine the impact of three key parameters, $M$, $\alpha$, and $\tau$, on the performance of the proposed algorithm. All analyses were performed under the simulation setting of Scenario 4 described in Section \ref{sec:simulation} in the supplementary materials.
	
\begin{figure}[!hbt]
		\centering
		\includegraphics[width=.9\textwidth]{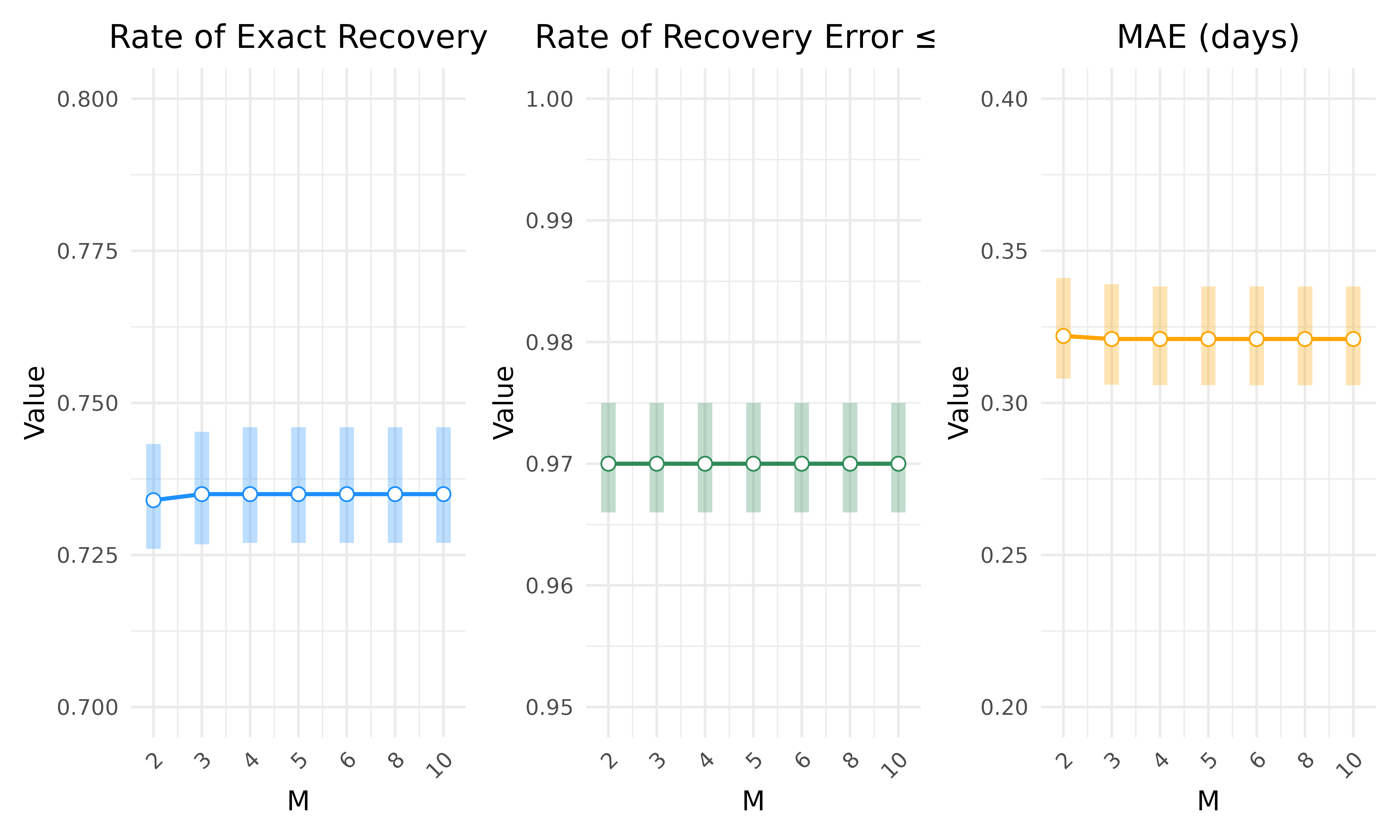}
		\caption[SA M.]{Sensitivity analysis on parameter $M$. The analysis is conducted under the simulation setting of Scenario 4; in each panel, the shaded band denotes the interquartile range (25th–75th percentiles) of the performance metric, and the central dot marks the median.}
		\label{fig:SAM}
	\end{figure}

	Parameter $M$ represents the maximum number of candidate clusters considered after dimensionality reduction. Smaller values of $M$ restrict the search space, while larger values (even beyond the true number of subtypes) preserve performance but increase computational cost. When users have prior biomedical knowledge about the number of disease subtypes, we recommend choosing $M$ slightly above that expectation. Sensitivity analysis over values of $M$ ranging from 2 to 10 demonstrated that performance remains stable and optimal when $M$ is equal to or greater than the true number of subtypes (four), supporting our recommendation (Figure~\ref{fig:SAM}).
	
\begin{figure}[!hbt]
		\centering
		\includegraphics[width=.9\textwidth]{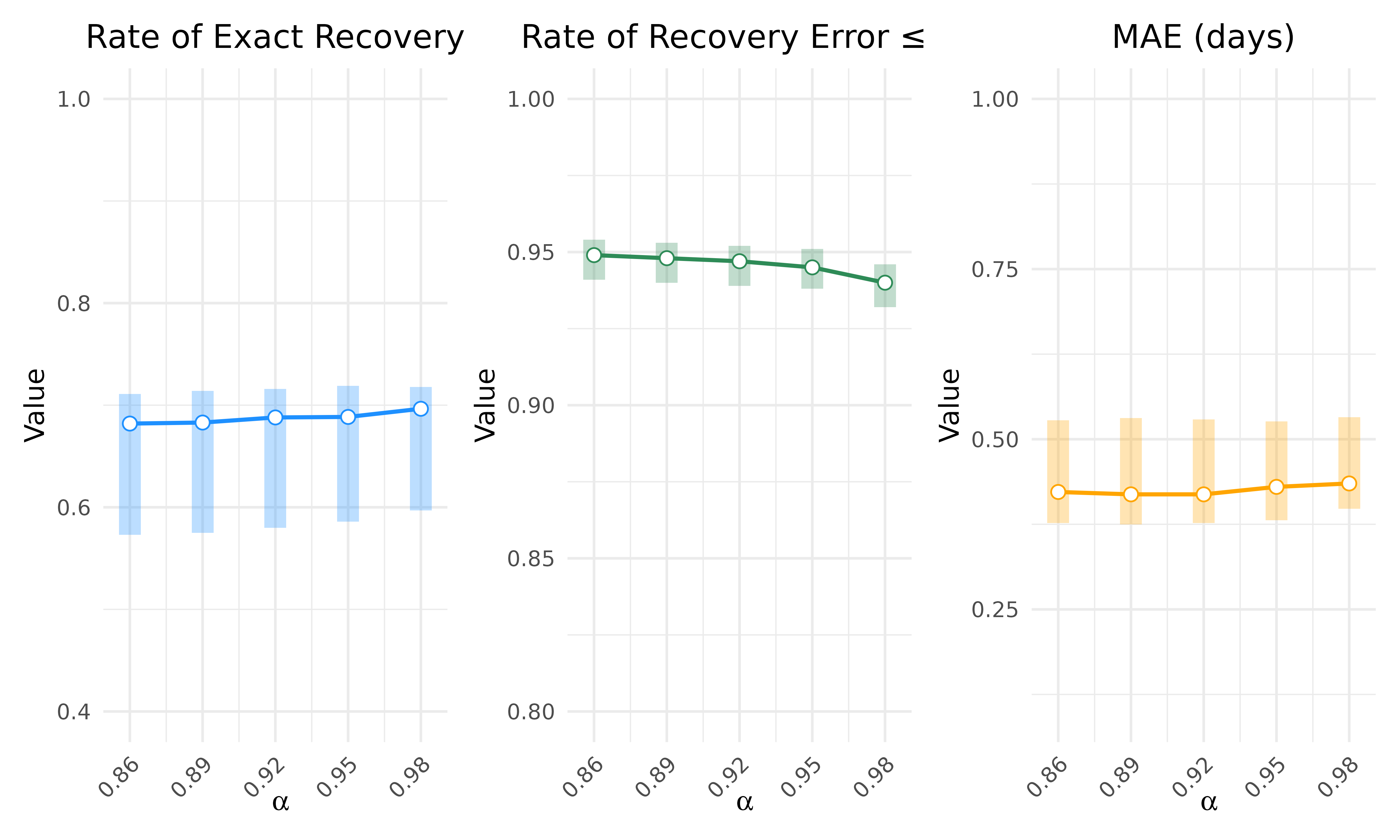}
		\caption[SA alpha.]{Sensitivity analysis on parameter $\alpha$. The analysis is conducted under the simulation setting of Scenario 4; in each panel, the shaded band denotes the interquartile range (25th–75th percentiles) of the performance metric, and the central dot marks the median.}
		\label{fig:SAalpha}
	\end{figure}

	Parameter $\alpha$ controls the robustness of average trajectory estimates within each cluster by masking potential outliers during timeline alignment. While selecting $\alpha$ is challenging without prior knowledge of contamination, recent theoretical results under Huber’s contamination model suggest that fully adaptive methods may incur substantial statistical cost \cite{luo2024adaptive}. Practically, $\alpha$ should not be set too high, as it may exclude too many data points and distort the representative trajectory. Sensitivity analysis across the range $[0.80, 0.99]$ confirmed that the algorithm’s performance is robust to this parameter (Figure~\ref{fig:SAalpha}).
	
\begin{figure}[!hbt]
		\centering
		\includegraphics[width=.9\textwidth]{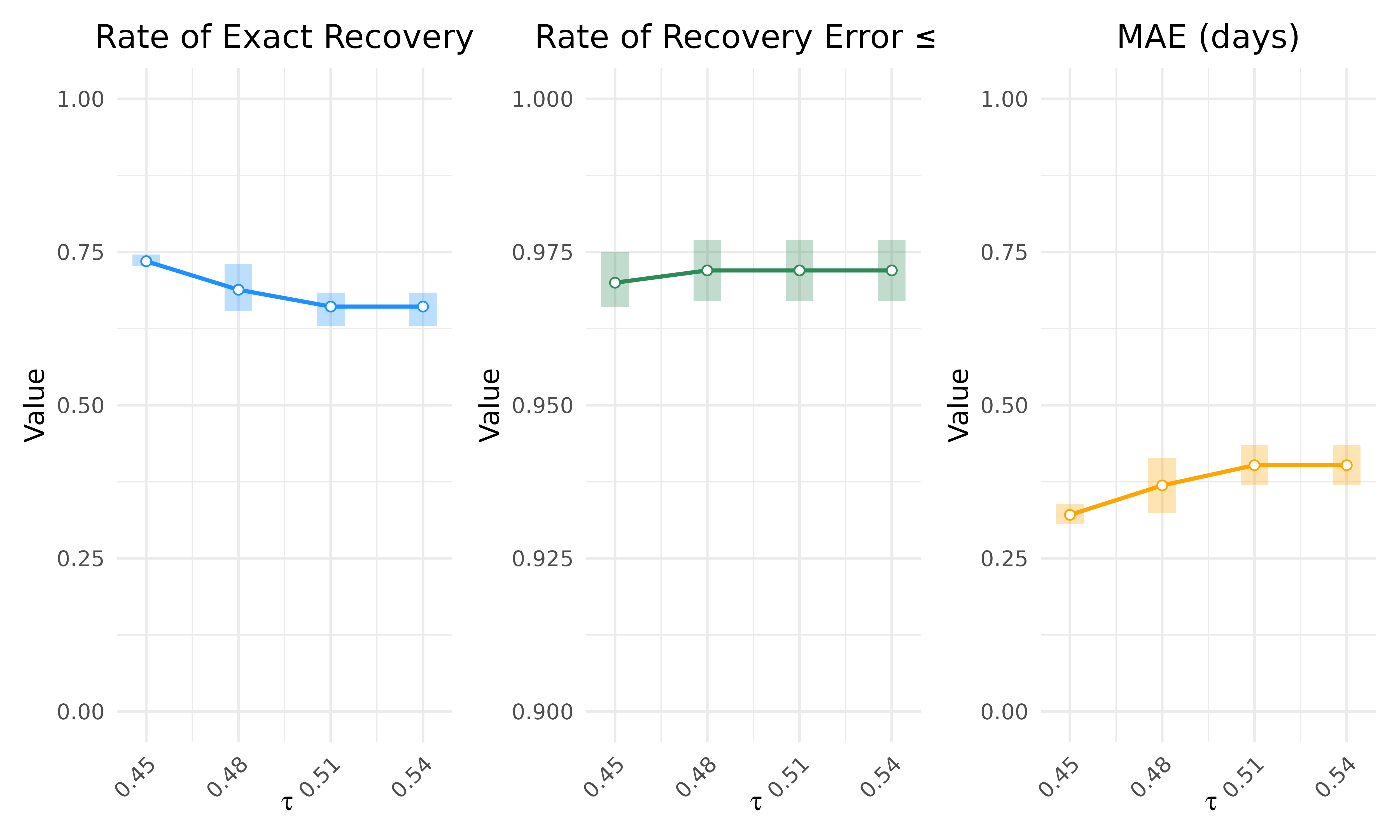}
		\caption[SA tau.]{Sensitivity analysis on parameter $\tau$. The analysis is conducted under the simulation setting of Scenario 4; in each panel, the shaded band denotes the interquartile range (25th–75th percentiles) of the performance metric, and the central dot marks the median.}
		\label{fig:SAtau}
	\end{figure}

	Parameter $\tau$ serves as a threshold for accepting the initial clustering, with higher values implying stronger confidence and fewer required iterations. In practical applications, users may refer to previously published recommendations, especially when using common public datasets. Additionally, general guidelines from the literature \cite{dalmaijer2022statistical} suggest interpreting silhouette coefficients greater than $0.5$ as evidence for clustering and coefficients exceeding $0.7$ as strong evidence. However, silhouette coefficients exceeding $0.7$ are rarely observed in biomedical applications \cite{banerjee2023identifying,manzini2022longitudinal}; coefficients around $0.5$ typically indicate sufficient quality. We conducted a sensitivity analysis over $\tau \in [0.45, 0.55]$ and observed stable algorithmic performance across this range (Figure~\ref{fig:SAtau}).
	
	\subsection{Ablation Studies} \label{sec:AS}

	\begin{table}[!ht]
		\centering
		\resizebox{\textwidth}{!}{%
			\begin{tabular}{c|c|c|c|c}
				\hline
				\textbf{Scenario} & \textbf{Rate of Exact Recovery} & \textbf{Rate of Recovery Error $\leq$ 1} & \textbf{MAE (days)} & \textbf{Runtime (min)} \\ \hline
				1 & 66.8\% [62.5\%, 74.8\%] & 78.8\% [72.8\%, 87.4\%] & 0.62 [0.42, 0.79] & 2.18 [0.92, 5.69] \\ \hline
				2 & 64.6\% [62.4\%, 68.3\%] & 87.4\% [77.2\%, 91.6\%] & 0.54 [0.47, 0.73] & 0.63 [0.50, 1.03] \\ \hline
				3 & 67.5\% [63.9\%, 72.0\%] & 77.4\% [72.5\%, 85.4\%] & 0.66 [0.50, 0.79] & 1.86 [0.98, 4.49] \\ \hline
				4 & 67.1\% [65.2\%, 69.0\%] & 87.3\% [85.3\%, 89.6\%] & 0.52 [0.48, 0.57] & 0.63 [0.50, 1.03] \\ \hline
				5 & 68.1\% [65.8\%, 70.6\%] & 82.4\% [75.8\%, 88.2\%] & 0.63 [0.51, 0.77] & 0.67 [0.51, 1.07] \\ \hline
				6 & 67.1\% [63.5\%, 71.7\%] & 77.5\% [73.6\%, 84.6\%] & 0.66 [0.52, 0.76] & 2.05 [1.01, 4.45] \\ \hline
				7 & 61.6\% [59.5\%, 64.3\%] & 82.6\% [79.9\%, 87.3\%] & 0.69 [0.57, 0.76] & 0.60 [0.53, 0.70] \\ \hline
				8 & 66.3\% [62.8\%, 72.1\%] & 78.3\% [73.1\%, 85.0\%] & 0.68 [0.53, 0.77] & 2.37 [0.97, 5.55] \\ \hline
			\end{tabular}%
		}
		\caption{Ablation Studies: We present the average value and 95\% confidence interval for the rate of exact recovery, rate of recovery error, mean absolute value, and runtime using the \textbf{smoothing spline} method.}
		\label{tab:Ablation Study1}
	\end{table}
	
	\begin{table}[!ht]
		\centering
		\resizebox{\textwidth}{!}{%
			\begin{tabular}{c|c|c|c|c}
				\hline
				\textbf{Scenario} & \textbf{Rate of Exact Recovery} & \textbf{Rate of Recovery Error $\leq$ 1} & \textbf{MAE (days)} & \textbf{Runtime (min)} \\ \hline
				1 & 55.7\% [52.7\%, 58.6\%] & 74.8\% [71.6\%, 79.2\%] & 0.88 [0.78, 0.96] & 1.09 [0.67, 1.57] \\ \hline
				2 & 49.9\% [46.4\%, 55.9\%] & 72.8\% [69.4\%, 77.6\%] & 0.97 [0.84, 1.06] & 1.16 [0.70, 1.73] \\ \hline
				3 & 53.4\% [50.4\%, 56.6\%] & 73.7\% [70.3\%, 78.3\%] & 0.93 [0.81, 1.02] & 1.09 [0.69, 1.57] \\ \hline
				4 & 50.2\% [47.3\%, 53.0\%] & 72.1\% [68.1\%, 76.2\%] & 0.98 [0.88, 1.09] & 1.10 [0.70, 1.56] \\ \hline
				5 & 53.3\% [50.8\%, 56.1\%] & 72.5\% [68.9\%, 76.8\%] & 0.95 [0.84, 1.04] & 0.88 [0.77, 1.01] \\ \hline
				6 & 54.4\% [51.9\%, 57.0\%] & 73.5\% [70.1\%, 77.8\%] & 0.92 [0.82, 1.01] & 0.87 [0.76, 1.01] \\ \hline
				7 & 46.1\% [40.3\%, 50.1\%] & 70.5\% [66.3\%, 73.8\%] & 1.06 [0.97, 1.18] & 0.89 [0.79, 1.06] \\ \hline
				8 & 53.2\% [50.6\%, 56.0\%] & 73.3\% [69.8\%, 77.8\%] & 0.94 [0.83, 1.02] & 0.87 [0.76, 1.05] \\ \hline
			\end{tabular}%
		}
		\caption{Ablation Studies: We present the average value and 95\% confidence interval for the rate of exact recovery, rate of recovery error, mean absolute value, and runtime using the \textbf{RKHS} (Reproducing Kernel Hilbert Space) method.}
		\label{tab:Ablation Study2}
	\end{table}
	
	\begin{table}[!ht]
		\centering
		\resizebox{\textwidth}{!}{%
			\begin{tabular}{c|c|c|c|c}
				\hline
				\textbf{Scenario} & \textbf{Rate of Exact Recovery} & \textbf{Rate of Recovery Error $\leq$ 1} & \textbf{MAE (days)} & \textbf{Runtime (min)} \\ \hline
				1 & 66.0\% [62.3\%, 86.6\%] & 82.0\% [76.9\%, 99.0\%] & 0.58 [0.15, 0.69] & 1.00 [0.52, 1.63] \\ \hline
				2 & 72.8\% [69.0\%, 77.0\%] & 91.9\% [88.3\%, 95.0\%] & 0.41 [0.32, 0.52] & 0.96 [0.52, 1.17] \\ \hline
				3 & 73.3\% [15.2\%, 78.7\%] & 88.2\% [61.9\%, 91.1\%] & 0.45 [0.36, 1.46] & 0.94 [0.52, 1.17] \\ \hline
				4 & 73.5\% [29.7\%, 80.2\%] & 93.4\% [80.0\%, 96.9\%] & 0.39 [0.31, 1.05] & 0.95 [0.52, 1.17] \\ \hline
				5 & 70.4\% [19.8\%, 77.9\%] & 86.6\% [62.4\%, 92.6\%] & 0.58 [0.43, 1.34] & 1.25 [0.52, 4.00] \\ \hline
				6 & 77.2\% [33.0\%, 81.3\%] & 93.8\% [85.3\%, 95.6\%] & 0.38 [0.31, 1.00] & 0.94 [0.52, 1.16] \\ \hline
				7 & 70.6\% [67.5\%, 74.4\%] & 91.7\% [90.5\%, 93.8\%] & 0.47 [0.39, 0.53] & 0.95 [0.51, 1.17] \\ \hline
				8 & 69.9\% [17.1\%, 75.1\%] & 85.5\% [48.0\%, 88.9\%] & 0.54 [0.43, 1.87] & 0.94 [0.51, 1.18] \\ \hline
			\end{tabular}%
		}
		\caption{Ablation Studies: We present the average value and 95\% confidence interval for the rate of exact recovery, rate of recovery error, mean absolute value, and runtime using the \textbf{k-means} method.}
		\label{tab:Ablation Study3}
	\end{table}
	
	To assess the dimension reduction component, we fixed the clustering method (using $k$-medoids) and compared our primary cubic B-spline dimension reduction approach (Table \ref{tab:Simulation_Results}) against two alternative dimension reduction methods: smoothing splines (Table \ref{tab:Ablation Study1}) and an approach based on Reproducing Kernel Hilbert Space (RKHS, Table \ref{tab:Ablation Study2}). All evaluated dimension reduction methods effectively recovered the true trajectories; the cubic B-spline approach consistently achieved the best overall performance, followed by the smoothing spline, and finally the RKHS-based method. These comparisons provide empirical justification for selecting cubic B-splines as our primary dimension reduction method.
	
	To evaluate the clustering component separately, we fixed the dimension reduction method (cubic B-spline) and compared the performance of two clustering algorithms: $k$-medoids clustering (Table \ref{tab:Simulation_Results}) and $k$-means clustering (Table \ref{tab:Ablation Study3}). Both clustering methods yielded sufficiently good results, though $k$-medoids generally outperformed $k$-means. Further comparisons between these clustering methods were conducted in the real-data analyses, and the results are summarized in Tables~1–5 of the main text. Both clustering algorithms were effective; but $k$-medoids clustering exhibited greater stability in unsupervised tasks (Sections~4.2) and delivered superior performance in clinical evaluation and supervised prediction tasks (Sections~4.3 and~4.4). The greater robustness of the $k$-medoids algorithm likely contributes to this observed advantage, providing practical guidance for clustering algorithm selection in practice.

	\section{Comparison with DTW Results} \label{appendix:CDTW}
	
	In this section, we compare the performance of our proposed registration method with the conventional dynamic time warping (DTW). Using the simulation setup described in Section~\ref{sec:simusetup}, we instantiate a scenario in which subjects exhibit heterogeneous disease progression speeds across subgroups. We consider two trajectory families: (i) a continuously heterogeneous family representing an imperfect subtyping setting, where some individuals are fast progressors (Group 2), some are slow progressors (Group 3), and the remaining display a continuum of intermediate speeds (Group 4); and
	(ii) a shape-heterogeneous family (Group 1), which shares a similar overall trend but differs in trajectory shape, again yielding imperfect subtyping relative to Groups 2–4.
	
	The data generation setup is as follows. Recall that $s_i^{(\cdot)} t_{ij}$ denotes the observed time, and $s_i^{(\cdot)}$ represents the subject-specific progression speed factor:
	\begin{align*}
		\text{Group 1 (300 subjects):}&\quad
		Y_i^{(1)}(t_{ij}) = 7 + 1.5\,t_{ij} + 4\sin\!\big(0.7(t_{ij}+4)\big) + \epsilon_{ij},\\[4pt]
		\text{Group 2 (250 subjects):}&\quad
		Y_i^{(2a)}\!\big(s_i^{(2)} t_{ij}\big) = 2\,t_{ij} + 3\sin(t_{ij}+4) + \epsilon_{ij},\\[4pt]
		\text{Group 3 (250 subjects):}&\quad Y_i^{(2b)}\!\big(s_i^{(3)} t_{ij}\big) = 2\,t_{ij} + 3\sin(t_{ij}+4) + \epsilon_{ij},\\[4pt]
		\text{Group 4 (200 subjects):}&\quad Y_i^{(2c)}\!\big(s_i^{(4)} t_{ij}\big) = 2\,t_{ij} + 3\sin(t_{ij}+4) + \epsilon_{ij}.
	\end{align*}
	where $s_i^{(2)} = 1$ for Group~2, $s_i^{(3)} =0.7$ for Group~3, and $s_i^{(4)} \sim \mathrm{Uniform}(0.7,1)$.
	
	We evaluate three choices of the number of clusters, $K_{ture} \in \{2,3,4\}$, with the ground-truth subtypes defined as:
	\begin{itemize}
		\item $K_{ture} =2$: Group~1 versus Groups~2--4, corresponding to two major subtypes with different shapes.\
		
		\item $K_{ture} =3$: Group~1 versus a relatively slow-progressor subtype versus a relatively fast-progressor subtype; the latter two constitute imperfect subtypes.\
		
		\item $K_{ture} =4$: Group~1, Group~2, Group~3, and Group~4, where Group~4 serves as a transitional subtype between Groups~2 and~3, again reflecting an imperfect subtyping configuration.
	\end{itemize}
	We then apply DTW and our registration method for downstream clustering and assess their ability to recover the true subtypes.\

	To evaluate clustering performance, we conducted 500 simulation replicates and computed three standard metrics: Adjusted Rand Index (ARI), Adjusted Mutual Information (AMI), and Clustering Accuracy (ACC). To ensure comparability across methods, we used a common protocol: (i) all clustering tasks were performed via $k$-medoids on the respective distance matrices; (ii) similar to the filtering in Section~4.1, we excluded individuals with no observations in either of the time windows $[0,8]$ or $(8,13]$ prior to computing the clustering metrics. These metrics respectively measure pairwise agreement, information overlap, and label correspondence between the true and predicted clusters. The average values and 95\% confidence intervals for all three metrics over 500 runs across all replicates are summarized in Table \ref{tab:dtw_vs_our}.
	
	\begin{table}[!ht]
		\centering
		
		\begin{tabular}{cccc}
			\hline
			\textbf{$K_{ture}$} & \textbf{Metric} & \textbf{DTW} & \textbf{Our Method} \\
			\hline
			2 & ACC & 0.697 [0.613, 0.748] & 0.635 [0.615, 0.657] \\
			& AMI & 0.078 [0.027, 0.214] & 0.110 [0.083, 0.140] \\
			& ARI & 0.112 [0.044, 0.198] & 0.067 [0.045, 0.093] \\
			\hline
			3 & ACC & 0.439 [0.405, 0.487] & 0.744 [0.722, 0.768] \\
			& AMI & 0.098 [0.048, 0.167] & 0.357 [0.317, 0.399] \\
			& ARI & 0.079 [0.035, 0.146] & 0.404 [0.363, 0.450] \\
			\hline
			4 & ACC & 0.431 [0.313, 0.507] & 0.588 [0.563, 0.616] \\
			& AMI & 0.206 [0.060, 0.348] & 0.327 [0.290, 0.360] \\
			& ARI & 0.170 [0.028, 0.291] & 0.321 [0.293, 0.352] \\
			\hline
		\end{tabular}
		\caption{Comparison of DTW and our method across different $K_{ture}$ values (average value [95\% CI])}
		\label{tab:dtw_vs_our}
	\end{table}
	
	Table \ref{tab:dtw_vs_our} shows that when the number of clusters is small ($K_{ture} = 2$), the two methods perform comparably. However, as the number of clusters increases ($K_{ture} = 3$ and $K_{ture} = 4$), our proposed registration method substantially outperforms DTW, achieving higher AMI and ARI scores and more stable clustering accuracy across replicates.

	\section{Computational and Space Complexities}
	
	In this section, we analyze the computational complexity of the proposed framework by examining each step separately. 
	
	In Step~1 (Algorithm~1), each subject’s longitudinal trajectory is transformed into a set of basis coefficients via ridge regression for each candidate time shift in $\Lambda$. Incorporating $p$ basis functions, the ridge regression for each shift incurs a computational cost of $\mathcal{O}(T_i p^2 + p^3)$, where $T_i$ is the number of observations for subject $i$. Aggregating across all subjects and all shifts, the total complexity for Step~1 is approximately $\mathcal{O}(|\Lambda|(Tp^2 + Np^3))$, where $T = \sum_{i=1}^N T_i$ denotes the total number of observations.
	
	In Step~2 (Algorithm~2), an iterative clustering procedure with dynamic time shift updates is performed. In each outer iteration, the algorithm evaluates $M$ candidate clusterings using $k$-means, $k$-medoids, or a similar algorithm, and computes the silhouette score for each $k$. Assuming the clustering procedure converges in $I_c$ inner iterations, this step requires $\mathcal{O}(MNkpI_c + MN^2p)$ per outer iteration, where $p$ is the feature dimension. Additional operations such as centroid computation, trimmed averaging, and time shift updates for each subject together contribute $\mathcal{O}(Np + N|\Lambda|p)$ per iteration. With $I$ outer iterations, the total complexity for Step~2 is $\mathcal{O}(I(MNkpI_c + MN^2p + Np + N|\Lambda|p))$.
	
	In Step~3 (Algorithm~3), unselected points are reassigned based on an exhaustive search over candidate shifts and clusters. For each unselected subject, this search requires evaluating $\mathcal{O}(|\Lambda|K^{(t_{\text{final}})}p)$ distances, where $K^{(t_{\text{final}})}$ is the number of clusters at termination. Assuming a constant fraction of subjects require reassignment, the total complexity for Step~3 is $\mathcal{O}(N|\Lambda|K^{(t_{\text{final}})}p)$.
	
	Summing across all three steps, the dominant terms are $\mathcal{O}(N|\Lambda|Tp^2)$ for Step~1 (since the number of basis functions $p$ is typically small and treated as a constant, e.g., around 10), $\mathcal{O}(IMN^2p)$ for Step~2 (since the number of subjects $N$ dominates the number of clusters, clustering iterations, and candidate shifts $|\Lambda|$), and $\mathcal{O}(N|\Lambda|Kp)$ for Step~3. Therefore, the overall computational complexity of the framework is $\mathcal{O}(N|\Lambda|Tp^2 + IMN^2p)$, as the contribution from Step~3 is relatively minor when $T$ is much larger than $K$. Here, $N$ is the number of subjects, $T$ is the total number of observations, $p$ is the number of basis functions, $|\Lambda|$ is the number of candidate time shifts, $M$ is the maximum number of clusters considered, $K$ is the number of clusters selected at convergence, and $I$ is the number of outer iterations. The computational cost is dominated by the data transformation step when the number of time shifts or basis functions is large, and by the clustering step when the number of subjects is large.
	
	On the other hand, for a given number of clusters $M$, the registration algorithm proposed in \cite{jiang2023timeline} incurs a computational cost of $\mathcal{O}(INMT_{\text{total}}|\Lambda|)$, where \( T_{\text{total}} \) denotes the total number of observations across the entire time grid. In the context of EHR data, this can be as large as \( \left|\{t_{ik} \mid i = 1, \ldots, N;\ k = 1, \ldots, T_i\}\right| \). Furthermore, if we aim to determine the optimal number of clusters by using the silhouette index in practice, this introduces an additional computational cost of \( \mathcal{O}(MN^2 T_{\text{total}}) \). In summary, the proposed algorithm yields a substantially lower computational cost compared to the method introduced in \cite{jiang2023timeline}.
	
	The space complexity of the proposed framework is primarily driven by the storage of transformed features and intermediate clustering outputs. In Step~1 (Algorithm 1), for each subject $i$ and each candidate time shift $l \in \Lambda$, the algorithm computes and stores a vector of $p$ spline coefficients. This leads to a three-dimensional tensor $\mathbf{\Omega}$ of size $N \times |\Lambda| \times p$, contributing a space complexity of $\mathcal{O}(N|\Lambda|p)$. In Step~2, only the coefficients corresponding to the currently selected shift $\delta_i^{(t)}$ for each subject are used in clustering, so no additional large data structures are required beyond temporary arrays of size $\mathcal{O}(Np)$ for centroid and distance calculations. Step~3 similarly reuses the stored coefficient tensor and does not introduce significant new memory demands. Therefore, the overall space complexity of the algorithm is $\mathcal{O}(N|\Lambda|p)$, where linear in the number of subjects, candidate shifts, and basis coefficients. This cost can be significantly lower than the space complexity of the method in \cite{jiang2023timeline}, which is \( \mathcal{O}(N T_{\text{total}}) \).
	
	In summary, our analysis of computational and space costs demonstrates the feasibility of applying our method to large-scale EHR datasets.

\end{sloppypar}

\end{document}